\documentclass[fleqn,usenatbib]{mnras}
\usepackage{hyperref}
\usepackage{newtxtext,newtxmath}
\usepackage{booktabs}

\usepackage[T1]{fontenc}
\usepackage{xcolor}

\DeclareRobustCommand{\VAN}[3]{#2}
\let\VANthebibliography\thebibliography
\def\thebibliography{\DeclareRobustCommand{\VAN}[3]{##3}\VANthebibliography}

\newcommand{\codefont}[1]{{\texttt{#1}}}
\usepackage{graphicx}	
\usepackage{amsmath}	
\usepackage{float}
\usepackage{accents}
\usepackage{cancel}

\title[Uncertainties in gravitational redshifts]{Cluster gravitational redshifts: uncertainties and survey requirements}

\author[E. Tsaprazi et al.]{
Eleni Tsaprazi$^{1}$ \thanks{e.tsaprazi@imperial.ac.uk}, Giorgio F. Lesci$^{2,3}$, Federico Marulli$^{2,3,4}$, Alan F. Heavens$^{1}$, Enrico Maraboli$^{5}$
\\
$^{1}$Imperial Centre for Inference and Cosmology (ICIC) \& Astrophysics group, Department of Physics, Imperial College, Blackett Laboratory,\\ Prince Consort Road, London SW7 2AZ, UK\\
$^{2}$Dipartimento di Fisica e Astronomia “Augusto Righi” - Alma Mater Studiorum Universita di Bologna, via Piero Gobetti 93/2, I-40129 Bologna, Italy\\
$^{3}$INAF - Osservatorio di Astrofisica e Scienza dello Spazio di Bologna, via Piero Gobetti 93/3, I-40129,Bologna, Italy\\
$^{4}$INFN - Sezione di Bologna, viale Berti Pichat 6/2, I-40127 Bologna, Italy\\
$^{5}$Dipartimento di Fisica, Università degli Studi di Milano, Via Celoria 16, I-20133 Milano, Italy
}

\date{Accepted XXX. Received YYY; in original form ZZZ}

\pubyear{2025}

\begin{document}
\label{firstpage}
\pagerange{\pageref{firstpage}--\pageref{lastpage}}
\maketitle

\begin{abstract}
We investigate the impact of observational and modelling uncertainties in cluster gravitational redshifts as a probe of modified gravity using an end-to-end forecasting pipeline. We use a generative model to build a halo catalogue with $M_{500}\ge 3\times 10^{13}\,M_\odot$, populate haloes with member galaxies via a five-parameter halo occupation distribution (HOD), assign projected positions from radial density profiles, apply survey-like selections, and infer a linear rescaling of the gravitational potential, $\alpha_\mathrm{MG}$, to parameterise modifications to general relativity (GR). We vary redshift uncertainties, radial and mass–redshift completeness, member abundance, minimum mass and maximum redshift, as well as mis-specify the clusters density and velocity profiles, centres, and mass function. We find that the intracluster velocity dispersion sets an effective floor: improving redshift precision beyond $\sigma_z\sim 10^{-4}(1+z)$ brings no improvement in the precision of $\alpha_\mathrm{MG}$. Realistic redshift and mass cuts primarily remove low-mass haloes and have minimal impact on the $\alpha_\mathrm{MG}$ precision. In this setting, we find that shallow, narrower spectroscopic surveys are preferable to deep, wide photometric ones for precise constraints on $\alpha_{\rm MG}$. We further find that mis-centring can mimic significant departures from GR. Baryonic deviations from a Navarro-Frenk-White profile and velocity anisotropies do not introduce appreciable biases. In the high-S/N regime of upcoming surveys, accurate determination of cluster centres will be essential to avoid interpreting systematic effects as new physics. The Spectroscopic Stage-5 Experiment and the Wide Field Spectroscopic Telescope provide a promising route toward establishing gravitational redshifts as a valuable and complementary probe of modified gravity.
\end{abstract}




\section{Introduction}\label{sec:intro}
General relativity (GR) \citep{1915SPAW.......778E}, the gravity theory of our standard cosmological framework, has been highly successful in describing gravitational phenomena across a wide range of scales. However, the unknown physical nature of cosmic acceleration and dark matter, together with persistent cosmological tensions in current observations \citep{2021CQGra..38o3001D}, raises the question of whether modifications of GR are required to explain cosmological observations. Following the detection of GW170817 \citep[see]
[and references therein]{2017PhRvL.119p1101A}, which provided strong evidence against theories that predict a modified speed of gravitational waves \citep{2017PhRvL.119y1304E}, attention has increasingly shifted toward cosmological tests of gravity across different scales and environments, at the level of simulations \citep{2018MNRAS.481.2813G,2020MNRAS.497.1885H,2024MNRAS.527.7242S,2025A&A...695A.170B}, theory \citep{2018PhRvD..97f1501L,2025JCAP...09..047M,2025PDU....4801906G}, and observations \citep{2024MNRAS.532.3972L,2024MNRAS.534..349L,2025arXiv251205819V}.

Galaxy clusters provide a powerful laboratory for cosmological tests of gravity \citep{Pizzuti_2025_refractedgravity}. One such probe is the gravitational redshift of cluster member galaxies: photons propagating in the vicinity of clusters experience a wavelength shift determined by the cluster gravitational potential. Theories of gravity that modify the potential in this density and scale regime can therefore be constrained by measuring this shift \citep{1995A&A...301....6C,2011Natur.477..567W,10.1098/rsta.2011.0291}. Examples include the Dvali–Gabadadze–Porrati (DGP) braneworld model \citep{2000PhLB..485..208D} in its normal (nDGP) and self-accelerating (sDGP) branches, as well as $f(R)$ gravity \citep{2007PhRvD..76f4004H,2010RvMP...82..451S}. The effects of the latter two can be directly parameterised as effective rescalings of a cluster’s gravitational potential \citep{2010PhRvD..81j3002S}. 

In this work, we develop an end-to-end pipeline that consists of (i) the construction of mock galaxy cluster and member catalogues, assuming a radial density profile and a halo occupation distribution, (ii) the prediction of the corresponding gravitational redshift signal, and finally, (iii) the inference of a modified gravity parameter. Our analysis is designed to address two key questions in the context of Stage-IV data releases and upcoming surveys: how do different modelling assumptions affect the predicted signal, and how do survey configurations impact the constraining power of cluster gravitational redshifts on modified gravity? We further interpret our results with the aim of informing both the conceptual design of Stage-V surveys \citep{2024arXiv240305398M,2025arXiv250307923B} and the observational and modelling uncertainty requirements needed to fully exploit this probe.

Significant efforts have been devoted to modelling gravitational redshifts \citep[e.g.][]{1990LNP...360...29N,2004ApJ...607..164K,2011Natur.477..567W,2013MNRAS.434.3008C,2013PhRvD..88d3013Z,2013MNRAS.435.1278K,2017arXiv170905756S,2017MNRAS.468.1981C,2024PhRvD.110j3523C,2024JCAP...05..003C,2025JCAP...07..080D}, extracting its observational signatures \citep{2000ApJ...533L..93B,2011Natur.477..567W,2015MNRAS.448.1999J,2015PhRvL.114g1103S,2021MNRAS.503..669M,2023A&A...669A..29R}, forecasting its constraining power \citep{2019IJMPD..2850150X,2022MNRAS.511.2732S,2025arXiv251213221T} and performing alternative gravitational redshift tests \citep{2026arXiv260119861D}. In this work, we provide an in-depth investigation of the survey-strategy discussion initiated in \citet{2025arXiv251213221T}, with the aim of translating those results into a form that is directly usable by the community. This discussion is particularly timely in view of the existing and imminent data releases from Stage-IV surveys. In particular, Euclid Data Release~3 \citep{2025A&A...697A...1E} is expected to deliver of order $10^5$ clusters over $14\,000$~deg$^2$, the DESI Legacy Imaging Survey \citep{2024ApJS..272...39W} covers $ 20\,000$~deg$^2$, and LSST will provide a deep cluster sample over $ 18\,000$~deg$^2$ \citep{2009arXiv09120201L}.

The paper is structured as follows.  In Sec. \ref{sec:model} we describe the theoretical model we will use for the prediction of the stacked gravitational redshift signal. In Sec. \ref{sec:method} we describe our mock generation pipeline. In Sec. \ref{sec:results} we present the main results of our analysis, with an emphasis on the impact of modelling choices and survey strategy on the gravitational redshift signal and its constraining power on modified gravity. Finally, in Sec. \ref{sec:conclusions} we summarise our findings and discuss how they can be used to inform upcoming analyses and the design of future surveys.

\section{Theoretical model}\label{sec:model}
\label{sec:model}

\begin{figure}
\centering
\includegraphics[width=0.455\textwidth]{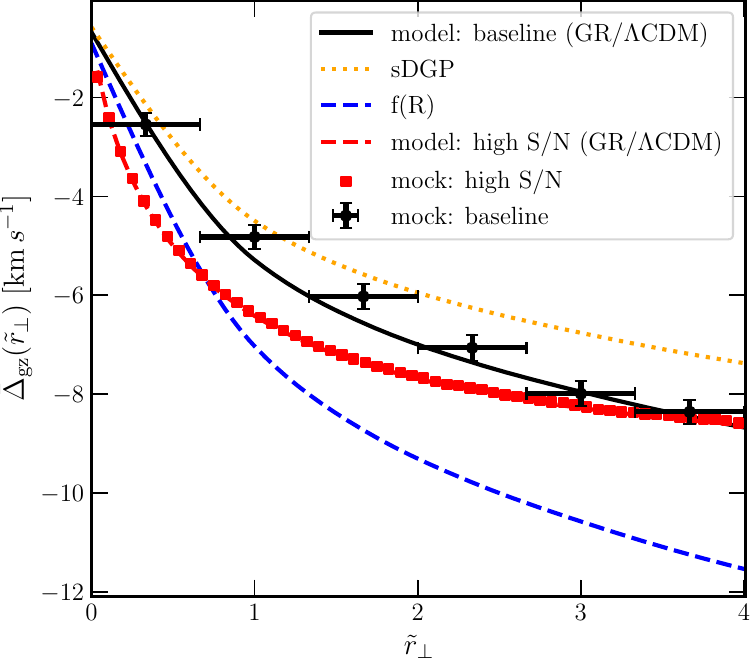}
\vspace{-0.5mm}
\caption{Model vs mock gravitational redshift in GR ($\alpha_{\rm MG}=1$) in the limit of no velocity dispersion (red) and the baseline case (black) with $1\sigma$ error bars. The sDGP and $f(R)$ predictions are shown in orange (dotted) and blue (dashed), respectively. We verify that the mock matches the model perfectly by artificially removing the velocity dispersions, and hence the transverse Doppler component, after which we recover perfect agreement in the limit of vanishing uncertainties.}
\label{fig_0}
\end{figure}

\begin{figure}
\centering
\includegraphics[width=0.47\textwidth]{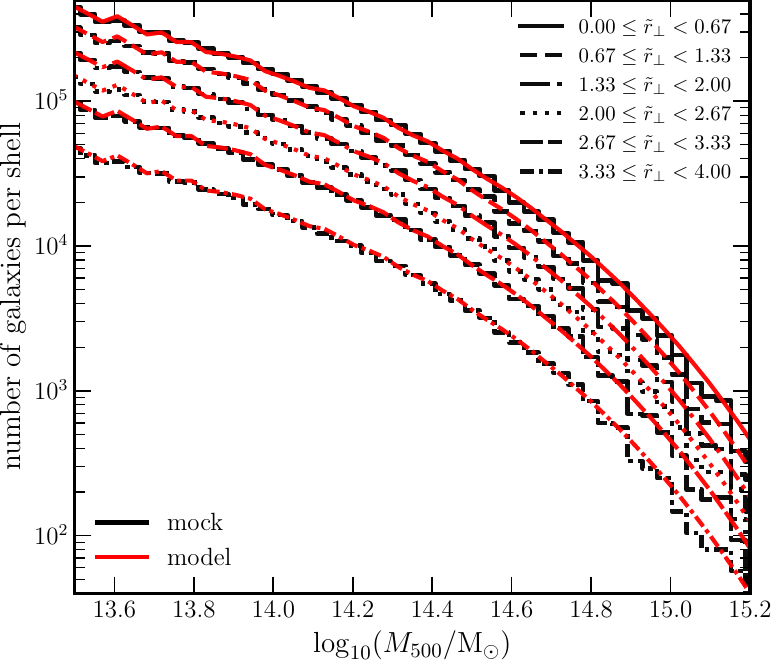}
\vspace{-1mm}
\caption{Predicted (red) vs mock (black) number of stacked galaxies per radial distance shell from the centre of the clusters as a function of halo mass in GR $\alpha_{\rm MG}=1$. This is the number of galaxies used for the estimation of the galaxy-weighted stacked signal. The match between the mock and the model indicate consistent stacking between the two.}
\label{fig_0b}
\end{figure}

In this section we describe the theoretical prediction for the stacked gravitational redshift signal that we will compare the mock measurements to. The observable we model is the gravitational redshift profile as a function of the normalised projected radius $\tilde{r}_\perp=r_\perp/r_{500}$ for a one-component, spherically symmetric Navarro-Frenk-White (NFW) total mass profile \citep{1997ApJ...490..493N}. Galaxy clusters may show more complex total mass structures, with multiple dark matter haloes which are not spherically symmetric \citep{Bonamigo_2018_RXCJ2248_M0416_M1206, Bergamini_2023_A2744SL}. Nevertheless, \citet{Maraboli_2025_virialquantitiesfromSL} showed that one-component, spherically symmetric total mass models accurately reproduce the total mass profiles of galaxy clusters even in the presence of complex substructures. 

For a halo with mass $M_{500}$, radius $r_{500}$, and concentration $c_{500}$, we define a generic dimensionless signal profile
\begin{equation}
  \bar{\Delta}_{\rm s,comp}(\tilde{r}_\perp)
  =
  \frac{2r_{500}}{\,\Sigma(\tilde{r}_\perp)}
  \times
    \int_{\tilde{r}_\perp}^{\tilde{r}_{\max}} f_{\rm comp}(\tilde{r})
    \frac{\,
\rho(\tilde{r})\,\tilde{r}\,\mathrm{d}\tilde{r}}
         {\sqrt{\tilde{r}^2 - \tilde{r}_\perp^2}},
  \label{eq:Delta_gz_single}
\end{equation}
where "${\rm comp}$" denotes the relativistic contribution to the single-halo signal, with ${\rm comp} \in \{{\rm gz}, {\rm TD}, {\rm LC}, {\rm SB}\}$ corresponding to the gravitational redshift, transverse Doppler, light-cone and surface brightness modulation terms, respectively. We discuss the latter in App. \ref{app:SB}. $M_{500}$ denotes the mass enclosed within $r_{500}$, inside which the mean density equals $500$ times the critical density of the Universe at the halo redshift and $\rho(\tilde{r})$ is the total mass density profile modelled with the NFW prescription
\begin{equation}
  \rho(\tilde{r}) =
  \frac{M_{500}\,c_{500}^{2}\,g(c_{500})}
       {4\pi r_{500}^{3}\,\tilde{r}\,[1+c_{500}\tilde{r}]^{2}}\;,
  \label{eq:nfw_rho}
\end{equation}
where the concentration parameter $c_{500}$ is given by \citet{2009ApJ...696.2115I},
\begin{equation}
  g(c_{500})^{-1} \equiv \ln(1+c_{500}) - \frac{c_{500}}{1+c_{500}}\;,
\end{equation}
and the projected surface mass density is
\begin{equation}
  \Sigma(\tilde{r}_\perp)
  = 2r_\mathrm{500}\!\int_{\tilde{r}_\perp}^{\tilde{r}_{\rm max}}
     \frac{\rho(\tilde{r})\,\tilde{r}\,\mathrm{d}\tilde{r}}{\sqrt{\tilde{r}^{2}-\tilde{r}_\perp^{2}}}\;,
  \label{eq:sigma}
\end{equation}
$\tilde{r}_{\max}$ is the maximum distance under consideration and $f_{\rm comp}(\tilde{r})$ is the function that determines which contribution we refer to\footnote{Our halo mocks come from a generative model rather than a lightcone, so we omit the corresponding term, which does not affect the generality of our conclusions. We further refer the reader to \citet{2025JCAP...07..080D}, who showed that the light-cone term is in general not present.} 

For the former component, $\bar{\Delta}_{\rm s,gz}$, we define
\begin{equation}
f_{\rm gz}(\tilde{r}) = \frac{\phi_0 - \phi(\tilde{r})}{c},
\label{eq:fgz}
\end{equation}
where $c$ is the speed of light and the gravitational potential is given by
\begin{equation}
\phi(\tilde{r}) = -g(c_{500})\frac{GM_{500}}{r_{500}}\frac{\ln(1+c_{500}\tilde{r})}{\tilde{r}}\,.
\label{eq:phi_potential}
\end{equation}
For the TD component, we define
\begin{equation}
f_{\rm TD}(\tilde{r}) = Q\frac{\tilde{r}^2-\tilde{r}_\perp^2}{c\tilde{r}}\frac{d\phi(\tilde{r})}{d\tilde{r}},
\label{eq:fTD}
\end{equation}
where we set $Q=3/2$ to assume isotropic galaxy orbits \citep{2013PhRvD..88d3013Z}. The above formulations dictate precisely how the mock cluster signal for the gravitational redshift and TD effect should be derived for a self-consistent mock generation. 

Modifications to gravity affect the clusters' gravitational potentials for a given intrinsic mass, leaving a signature to the single-halo gravitational redshift profile primarily through Eq. \eqref{eq:fgz}. Specifically, as we discuss below, given a cluster with the same intrinsic mass, in f(R) and sDGP gravity, the potential will be deeper and shallower than in general relativity, respectively. In this work, we follow the modified gravity parameterisation adopted by \citet{2023A&A...669A..29R}. According to this prescription, the potential $\phi(\tilde{r})$ depends linearly on the amplitude parameter $a_\mathrm{MG}$ through which deviations from GR are encoded as an effective gravitational constant, $G_\mathrm{eff}$, linearly rescaled with respect to the gravitational constant in GR, such that $G_\mathrm{eff}=a_\mathrm{MG}G$. In GR, $a_\mathrm{MG}=1$. \citet[][Eq. 29]{2010PhRvD..81j3002S} showed that in $f(R)$ gravity, where the Einstein-Hilbert Lagrangian is $[R+f(R)]/16\pi G$, gravitational forces are enhanced by $a_\mathrm{MG}=4/3$ for a scalar field obeying $|f_{R,0}|=10^{-4}$ in the strong-field case\footnote{$|f_{R,0}|=10^{-6}$ produces screened haloes, while $|f_{R,0}|=10^{-5}$ constitutes an intermediate scenario.}. Further, \citet[][Eq. 19]{2023A&A...669A..29R} use $a_\mathrm{MG}\approx0.85$ to mimic the \citet{2010PhRvD..81j3002S} setting in the sDGP scenario. The latter case corresponds to a self-accelerating model without a cosmological constant term, calibrated to match Cosmic Microwave Background observations and cosmic expansion data. In this model, the gravitational potential depends on the ratio of the distance from the cluster centre to the Vainshtein radius, $r_{*}$, which is itself determined by the mass enclosed within that radius. Modified gravity effects are suppressed at distances $\ll r_{*}$, while on larger scales the modified gravity parameter approaches the aforementioned asymptotic value. For the full mathematical formulation of this model, we refer the reader to \citet[Eqs.~47--52]{2010PhRvD..81j3002S}. Future work should investigate how specific viable modified-gravity theories alter the gravitational potentials of galaxy clusters \citep{2017PhRvL.119y1304E}. The modified gravity parameter, $\alpha_{\rm MG}$, is set to unity throughout our analysis, while the $f(R)$ and sDGP curves are shown only as phenomenological reference models.

Modified gravity in principle affects halo concentrations \citep{2019MNRAS.487.1410M,2021MNRAS.508.4140M}, halo masses \citep{2018MNRAS.477.1133M}, halo occupation distributions \citep{2018MNRAS.479.4824H}, halo mass functions (HMF) \citep{2016JCAP...12..024C,2017JCAP...03..012V,2024A&A...691A.323S}, velocity dispersions \citep{2012MNRAS.425.2128J} and radial density profile \citep{2024JCAP...11..014P}. Further, we note that Eq. \eqref{eq:phi_potential} holds in GR, because it relies on the Poisson equation relating the density profile to the potential. The potential causing gravitational redshift is the temporal distortion, while the potential that enters the Poisson equation is the spatial distortion. While these two potentials coincide in GR, they generally differ in modified gravity theories. Analyses aiming to constrain specific modified-gravity theories, rather than simply quantify deviations from the GR case ($\alpha_{\rm MG}=1$) as done here, should vary the modelling assumptions self-consistently when deriving the theoretical predictions. Therefore, the $f(R)$ and sDGP reference models shown here are obtained by a simple rescaling of the GR prediction, following \citet{2023A&A...669A..29R}. Since these reference curves are not used in the analysis, this choice does not affect our constraints. Nevertheless, its validity should be tested explicitly, in particular with modified-gravity simulations \citep[e.g.][]{2018MNRAS.481.2813G,2026arXiv260313148E}.

In the model, we evaluate all redshift-dependent quantities at a single pivot redshift, $z_\mathrm{piv}$, which we take to be the median of the cluster redshifts. We find that this choice does not shift the signal with respect to the theory prediction with respect to integrating over the redshift range considered. The mean or the mode produce more biased predictions which become noticeable close to $4r_{500}$. We find that this pivot redshift can adequately characterise the redshift distribution of the cluster population even in the limit of no velocity dispersion as shown in Fig. \ref{fig_0}. In the present study, we adopt this choice for fast model evaluation.

We assume the \citet{Tinker08} HMF and include the mass--redshift completeness factor $C_{\rm cl}(M,z)$, both evaluated at $z_{\rm piv}$ and denoted $C_{\rm cl}(M,z_{\rm piv})$. This function characterises the quality of the cluster catalogue which would be produced by a cluster finder. We refer the reader to \citet{2019A&A...627A..23E} for an in-depth discussion of related systematic effects, the modelling of which, we will postpone, however, to a future analysis aimed at full survey realism. At fixed projected radius $\tilde{r}_\perp$ we then define the mass-weighted, surface-density--weighted stack of the total profile as
\begin{equation}
  \bar{\Delta}_{{\rm comp}}(\tilde{r}_\perp)
  =
  \frac{
    \displaystyle
    \int_{M_{\rm min}}^{M_{\rm max}} \! {\rm d}M\,
    \mathcal{W}\,
    \bar{\Delta}_{\rm s,comp}(\tilde{r}_\perp)
  }{
    \displaystyle
    \int_{M_{\rm min}}^{M_{\rm max}} \! {\rm d}M\,
    \mathcal{W}}\;,
    \label{eq:Delta_stack_mass}
\end{equation}
where the weights are defined as
\begin{equation}
\mathcal{W}=\Phi(M,z_{\rm piv})\,
    \langle N_{\rm HOD}(M)\rangle\,
    C_{\rm cl}(M,z_{\rm piv})\,
\Sigma(\tilde{r}_\perp)\;,
\label{eq:weighting}
\end{equation}
where $\Phi$ is the HMF and $N_{\rm HOD}(M)$ is the halo occupation distribution (HOD) of both central and satellite galaxies, described below. This formulation assumes that the HMF above is the one of the sample with perfect halo and member galaxy completeness. The total signal, $\bar{\Delta}$, is the sum of the "${\rm comp}$" components in Eq. \eqref{eq:Delta_stack_mass}, such that $\bar{\Delta}=\bar{\Delta}_{\rm gz}+\bar{\Delta}_{\rm TD}$ for our main analysis. We include the effect of surface brightness modulation only in the flux-limited survey forecasts discussed in Sec. \ref{sec:SB} and describe the implementation in App. \ref{app:SB}.

In practice, $\bar{\Delta}$ is obtained from the histogram of redshift differences between the cluster members and the centre and is therefore a measure of the shift of the mean of this distribution. Here, we work in the rest frame of the clusters. In the real Universe, the signal contains a contribution from evolution terms due to the background cosmic expansion. This can be removed by fitting the redshift difference distribution as the sum of a quasi-Gaussian distribution and a linear slope \citep{2011Natur.477..567W}. Weighting with $\langle N_{\rm HOD}(M)\rangle$ has a minor effect when a realistic cut of $M_\mathrm{min}>10^{14}\,M_\odot$ is assumed, but its impact becomes greater for lower mass cuts. We find that this weighting is necessary for the model predictions to accurately match the mocks in our setting, as shown in Fig. \ref{fig_0b}, which shows the number of galaxies used for the estimation of the signal per radial shell. Given that the mock and model predicted number of member galaxies match, we conclude that the model and mock stacking is consistent.

Our baseline analysis assumes that the galaxy population follows the five-parameter HOD of \citet{2005ApJ...633..791Z}. Assuming that the central and satellite galaxy halo occupation distributions are independent, we write their total halo occupation distribution as
\begin{equation}
  \langle N_{\rm HOD}(M) \rangle
  = \bigl\langle N_{\rm cen}(M)\bigr\rangle
    \bigl[1+\bigl\langle N_{\rm sat}(M)\bigr\rangle\bigr]\;,
\end{equation}
with a central term
\begin{equation}
  \bigl\langle N_{\rm cen}(M)\bigr\rangle
  = \frac{1}{2}\left[
      1+\mathrm{erf}\!\left(
        \frac{\log_{10}M-\log_{10}M_{\min}}
             {\sigma_{\log M}}
      \right)
    \right]\;,
\end{equation}
and a satellite term which yields
\begin{equation}
  \bigl\langle N_{\rm sat}(M)\bigr\rangle
  = \left(
      \frac{M-M_{\rm cut}}{M_{1}}
    \right)^{\!\alpha_0},
  \qquad M>M_{\rm cut}\;,
\end{equation}
and $\langle N_{\rm sat}(M)\rangle=0$ otherwise. For central galaxies, $M_\mathrm{min}$ is the minimum mass of haloes that can host central galaxies, while $\sigma_{\log M}$ modulates the width of the distribution. For satellite galaxies, $M_\mathrm{min}$ represents the truncation mass, $M_1$ the normalisation and $\alpha_0$ the power-law index. Here $r_{500}(M,z)$ is computed from 
\begin{equation}
  \begin{split}
  r_{500}(M, z)
  &=
  \left[
      \frac{3M}{4\pi\,500\,\rho_{\rm c}(z)}
    \right]^{1/3},
  \end{split}
  \label{eq:r2m}
\end{equation}
for $z=z_\mathrm{piv}$, using the critical density $\rho_{\rm c}(z)$ of the Planck~2018 cosmology. 

For the purposes of investigating the impact of mis-specifying the radial density profile, we will also consider a toy cuspy model that scales as
\begin{equation}
  \rho_\mathrm{bar}(\tilde{r})
  =
  \,\mathcal{K}\,
  \rho(\tilde{r};M_{500},c_{500},r_{500})
  \times\left[1 + f_\mathrm{bar}\,
    \exp\!\left(-\frac{\tilde{r}^{2}}{\tilde{r}_\mathrm{bar}^{2}}\right)\right]\;,
  \label{eq:nfw_bar}
\end{equation}
where $f_\mathrm{bar}$ controls the amplitude of the central baryonic compression and $\tilde{r}_\mathrm{bar}$ sets a dimensionless characteristic scale. We adopt $f_\mathrm{bar}=3$ and $\tilde{r}_\mathrm{bar}=0.05$ for a noticeable but not extreme compression, and $f_\mathrm{bar}=6$ for a heavier compression, as can be seen upon comparison of the profiles to \citet[][Fig. 2]{2022A&A...665A.143L}. The normalisation factor $\mathcal{K}$ is chosen such that the enclosed mass within $r_{500}$ remains equal to $M_{500}$, so that the toy profile preserves the total mass by construction while modifying only the inner density structure. 

While spherical symmetry may be a reasonable approximation for a large stacked cluster population, \citet{2017MNRAS.468.1981C} showed that the galaxy-weighted gravitational-redshift profile may be biased by the non-spherical morphology of individual haloes, neighbouring structures, and the surrounding cosmic web. To test the sensitivity of the signal to this effect, we introduce a toy asymmetry model in which the otherwise spherical satellite distribution is supplemented, for a subset of systems, by a correlated neighbouring NFW halo. This neighbouring component is intended to mimic the secondary potential minima and anisotropic galaxy weighting expected in realistic cluster environments. Unlike the spherical and baryon-compressed variants, this model is not a purely radial profile. It consists of a secondary halo that has $30\%$ of the mass of the primary one and is located outside the main cluster core, but within the few-\(r_{500}\) regime where correlated structures and secondary potential minima are expected to affect the galaxy-weighted gravitational-redshift signal.

In Fig.~\ref{fig_0} we show that in the limit of no peculiar velocities we recover perfect agreement between the model and the mocks, and we contrast this with our baseline configuration which assumes an ideal uncertainty scenario (variant~0 in Table~\ref{tab:variants}). 

\section{Mock catalogue construction}\label{sec:method}

We generate our mock halo catalogue using the generative model described below. As the first step, we assume a $\Lambda$ cold dark matter ($\Lambda$CDM) cosmological model, adopting the parameters from \citet[][Table 2, TT, TE + EE + lowE + lensing]{2020A&A...641A...6P}. Haloes span the redshift range $z\in[0.05,2]$ and the mass range $M_{500}\in[0.3,20]\times10^{14}$ M$_\odot$. The number of haloes in the catalogue, $N_{\rm h}$, is defined as follows:
\begin{equation}
N_{\rm h} = \Omega \int_{z_{\rm min}}^{z_{\rm max}}{\rm d}z\,\frac{{\rm d}^2 V}{{\rm d} z{\rm d}\Omega}\int_{M_\mathrm{min}}^{M_{\rm max}}{\rm d}M_{500}\,\frac{{\rm d} n(M_{500},z)}{{\rm d} M_{500}}\,,
\end{equation}
where $z_{\rm min}$ and $z_{\rm max}$ are the minimum and maximum redshifts, respectively, while $M_\mathrm{min}$ and $M_{\rm max}$ are the minimum and maximum halo masses, respectively. In addition, ${\rm d}\Omega$ is the solid angle element, $\Omega$ being the total survey area in steradians, corresponding to 18\,000 deg$^2$, $V$ is the comoving volume, while ${\rm d} n(M,z)/{\rm d} M$ is the HMF, for which we adopt the model by \citet{Tinker08}. We sample $N_{\rm h}$ pairs $(M_{500},z)$ from the joint probability density $P(M_{500},z)=P(z)P(M_{500}|z)$, where
\begin{equation}\label{eq:Pz}
P(z) = \frac{\Omega}{N_{\rm h}}\,\frac{{\rm d}^2 V}{{\rm d} z{\rm d}\Omega}\int_{M_\mathrm{min}}^{M_{\rm max}}{\rm d}M_{500}\,\frac{{\rm d} n(M_{500},z)}{{\rm d} M_{500}}\,,
\end{equation}
and
\begin{equation}\label{eq:PMz}
P(M_{500},\,z) = \frac{\Omega}{N_{\rm h}}\frac{{\rm d}^2V}{{\rm d}z{\rm d}\Omega}\,\frac{{\rm d} n(M_{500},z)}{{\rm d} M_{500}}\,.
\end{equation}
The $(M_{500},\,z)$ pairs are obtained through Monte Carlo sampling of Eq. \eqref{eq:PMz}. We derive the corresponding $r_{500}$ using Eq. \eqref{eq:r2m}. Subsequently, we populate the haloes with member galaxies generating as many cluster members per cluster as dictated by the HOD. We assume the following HOD parameters $\log_{10}(M_{\min}/M_\odot)=11.68$,
$\sigma_{\log M}=0.15$,
$\log_{10}(M_{\rm cut}/M_\odot)=11.86$,
$\log_{10}(M_{1}/M_\odot)=13.0$,
$\alpha_0=1.02$. This configuration  corresponds to about $10$ members per cluster at $M_\mathrm{500}=10^{14}$ M$_\odot$ \citep[Table 1,][]{2005ApJ...633..791Z}. For each halo the expected value is converted to an integer via
\begin{equation}
  N_{\rm HOD}
  = \bigl\lfloor
      \bigl\langle N_{\rm HOD}(M)\bigr\rangle
    \bigr\rfloor\;,
\end{equation}
where $\lfloor\cdot \rfloor$ indicates the floor operator. The above expression represents the mean number of galaxies per cluster. For each cluster we draw $N_{\rm HOD}$ 3D distances from the cluster centre from Eq. \eqref{eq:nfw_rho} and projected distances from Eq. \eqref{eq:sigma} per cluster. 

We then proceed to the modelling of observational systematic effects. Starting from the parent mock described above, we impose a radial galaxy completeness and a mass--redshift dependent halo completeness. For each cluster we first thin the satellite population according to a radial completeness function $C_{\rm r}(\tilde r_\perp)$ that decays with distance from the cluster centre, as observed in typical cluster studies \citep[][Fig. 3]{2013A&A...558A...1B}. For the purposes of this work, we consider the decaying profile $
C_{\rm r}(\tilde r_\perp) = C_1\exp\!\left(-C_2\,\tilde r_\perp\right),$ where $C_1,C_2$ modulate the shape of the completeness as a function of distance from the cluster's centre. This selection mimics a flux-based selection but in the space of distance rather than magnitudes. We will consider $C_1=0.9$ and $C_2=0.6$ to mimic the member completeness in a galaxy survey. This setting yields around $50\%$ probability of observation at $1r_{500}$ and around $10\%$ probability at $4r_{500}$. In our analysis variants, we will also consider the same profile but with a plateau close to the cluster centre ($C_1=0.9, C_2=0$ out to $1r_{500}$) to mimic the effect of a cluster-targeting survey with increased and constant completeness close to the cluster core, which can be achieved for example with integral field spectroscopy \citep{2026arXiv260215934M}. We will vary these values to assess the effect of varying number of galaxy members, as shown in Table \ref{tab:variants} (variants 5, 6). Each galaxy with projected radius $\tilde r_{\perp}$ is retained with probability $C_{\rm r}(\tilde r_{\perp})$ via rejection sampling. Galaxies that fail this test are removed from the mock. This leaves, for each cluster $i$, an observed richness $N_{\rm mem}(i)$ which is lower than the parent HOD prediction.

At the cluster level, we impose an additional completeness factor that depends on halo mass and redshift, $C_{\rm cl}(M_{500},z_{\rm c})$ which mimics a cluster-finder systematic effect. We model this empirically as a smoothed step in $\log_{10}M_{500}$ whose threshold increases with redshift via 
$\log_{10}M_{\rm cut}(z) = M_{\rm cut,0} + \alpha_\mathrm{cut} z,
$ and
\begin{equation}
  x(M,z) = \frac{\log_{10}M - \log_{10}M_{\rm cut,0}(z)}
                {\sigma_{\log M}}\;,
\end{equation}
with $(M_{\rm cut,0},\alpha_\mathrm{cut},\sigma_{\log M})$ being the free parameters. We will vary $M_{\rm cut,0}$ in the different variants entering our analysis and fix $\alpha_\mathrm{cut}=0.1$, $\sigma_{\log M}=0.32$. The cluster completeness is then
\begin{equation}
  C_{\rm cl}(M,z)
  = \frac{1}{1 + \exp\!\bigl[-2x(M,z)\bigr]}\;.
  \label{eq:cluster_completeness}
\end{equation}
This expression yields a smooth transition from $C_{\rm cl}\simeq 0$ for $M\ll M_{\rm cut,0}$ to $C_{\rm cl}\simeq 1$ for $M\gg M_{\rm cut,0}$, reproducing a smooth decrease with redshift and halo mass. For $M_{\rm cut,0}=13.8$, our choice of parameters produces a setting where a cluster of mass $10^{14}$ M$_\odot$ has $65\%$ and $50\%$ probability of being observed at $z=1$ and $z=2$, respectively. This setting represents a conservative choice based on \citet{2019A&A...627A..23E}. The halo completeness is applied via rejection sampling.

The gravitational redshift contribution for a galaxy is computed from Eq. \eqref{eq:fgz}. Subsequently, we generate sets of three-dimensional cluster member velocities. In the baseline isotropic case, the three-dimensional total velocity dispersion of cluster members is related to the TD kernel through
\begin{equation}
\sigma^2(\tilde r,\tilde r_\perp)=2\,f_{\rm TD}(\tilde r,\tilde r_\perp)\,.
\label{eq.veldispfromTD}
\end{equation}
This quantity is not the velocity dispersion measured from the width of the stacked redshift-difference distribution. Rather, it is the local three-dimensional velocity-dispersion scale assigned to a mock galaxy at fixed $(\tilde r,\tilde r_\perp)$, chosen so that the resulting transverse-Doppler contribution is consistent with Eq.~\eqref{eq:fTD}; the observable line-of-sight dispersion is obtained only after projection, stacking over the cluster population, and inclusion of redshift uncertainties. This follows from \citet[][Eq. 5]{2013PhRvD..88d3013Z} and assumes that in the rest-frame of our mock galaxy clusters the mean velocity is zero. When the velocity dispersion is isotropic,
\(\sigma_r^2=\sigma_\theta^2=\sigma_\phi^2=\sigma^2/3\), where \((r,\theta,\phi)\) denote spherical coordinates centred on the cluster.

To explore the impact of velocity anisotropy, we introduce the standard anisotropy parameter
\begin{equation}
\beta = 1-\frac{\sigma_\theta^2+\sigma_\phi^2}{2\sigma_r^2}.
\end{equation} 
Employing the TD prescription of Eq. \eqref{eq.veldispfromTD} and imposing $\sigma^2=\sigma_r^2+\sigma_\theta^2+\sigma_\phi^2$, we arrive at
\begin{equation}
    \sigma_r^2=\frac{\sigma^2}{3-2\beta}\,.
\end{equation}
By assuming symmetry between the two tangential directions, the tangential dispersions become 
\begin{equation}
\sigma_\theta^2=\sigma_\phi^2=(1-\beta)\sigma_r^2.
\end{equation}

The velocity components of each cluster member, $(v_r, v_\theta, v_\phi)$, with respect to the observer, are then drawn from Gaussian distributions with these dispersions \citep{Mamon_2013_MAMPOSSt,AguirreTagliaferro_2021_MAMPOSStvalidationNbodysim,Read_2021_MAMPOSStvalidationNbodysim}. The component projected along the line of sight (LOS) is then defined as
\begin{equation}
v_{\rm pec}=\mu v_r+\sqrt{1- \mu^2}\left(v_\theta\cos\varphi+v_\phi\sin\varphi\right),
\label{eq:velocity_model}
\end{equation}
where \(\varphi\) is an azimuthal angle drawn uniformly between \(0\) and \(2\pi\) which determines the orientation of the tangential velocity vector relative to the LOS projection direction. The geometric projection factor is $\mu=\pm\sqrt{1-\tilde r_\perp^2/\tilde r^2}$, with the sign drawn at random to account for the two possible LOS positions at fixed \((\tilde r,\tilde r_\perp)\). Because the above prescription depends explicitly on \(\tilde r_\perp\), the resulting velocity dispersion is not strictly isotropic. The above LOS component of our mock 3D velocities is the final input to our pipeline, from which we infer the LOS velocity dispersion and compute the mean velocity shift.

The TD kernel itself further assumes isotropy via \(Q=3/2\). The parameter \(\beta\) therefore introduces anisotropy at the level of the velocity decomposition, redistributing the velocity dispersion between radial and tangential components relative to the isotropic baseline. We will investigate the impact of anisotropy in two ways. First, we will keep \(Q=3/2\), but introduce a non-zero $\beta$. In the second case, we will modify $Q$, keeping $\beta=0$. The former should be understood as a phenomenological stress test, rather than a fully self-consistent anisotropic Jeans solution, since the value \(Q=3/2\) itself follows from the isotropic case. For the first test, we will assume constant $\beta=0.2$ and $\beta=-0.2$ following \citet[][Fig. 17]{2024MNRAS.533.3647M} and \citet{2026A&A...707A.153B}. An alternative way to break the assumption of isotropy is to keep \(\sigma_r^2=\sigma_\theta^2=\sigma_\phi^2=\sigma^2/3\), while varying $Q$ from the baseline $Q=3/2$ to $Q=(3r_{\rm vir}+r)/(2r_{\rm vir}+r)$  \citep{2013PhRvD..88d3013Z}, where $r_{\rm vir}$ is the virial radius of the cluster, keeping $\beta=0$. To mimic this scenario, we adopt the conversion $r_\mathrm{vir}=(c_\mathrm{vir}/c_{500})\,r_\mathrm{500}$ \citep{2003ApJ...584..702H, 2015MNRAS.450.3665S}, yielding approximately for $r_\mathrm{vir}\approx1.6r_\mathrm{500}$ for our baseline sample. The above 2D prescription is sufficient for the purposes of constructing a model for the present analysis, yet future velocity models should adopt the more computationally expensive, yet full 3D solution to the Jeans' equation per \citet[][Eq. 3]{2026A&A...707A.153B}, as anisotropy can be highly relevant for accurate dynamical modelling of galaxy clusters \citep{2026arXiv260215934M}.

The corresponding redshift contributions are
\begin{equation}
z_{\rm pec} = \frac{v_{\rm pec}}{c}\,,
\end{equation}
and
\begin{equation}
z_{\rm TD}  = \frac{v_r^2 + v_\theta^2 + v_\phi^2}{2c^2}\,,
\end{equation}
where $z_{\rm pec}$ is the standard Doppler shift arising from the line-of-sight peculiar velocity defined in Eq.~\eqref{eq:velocity_model}, and $z_{\rm TD}$ denotes the transverse Doppler contribution associated with the squared magnitude of the three-dimensional velocity. The radial velocity component has a dispersion of around $450$ km$\,s^{-1}$, measured from the mock catalogue, setting a noise threshold at $0.0015=450/c$ in redshift, beyond which we find minimal improvement in the $\bar{\Delta}$ error bars for fixed number of members.

We additionally include measurement uncertainties on the member redshifts. The redshift error is modelled as a Gaussian perturbation $z_{\rm err} \sim \mathcal{N}(0,\,\sigma_{z}^2(1+z)^2)$, with $\sigma_{z}$ set by the assumed redshift precision. The total redshift for each galaxy relative to the true halo centre is then
\begin{equation}
  (1+z_{\rm tot})
 = (1+z_{\rm gz})(1 + z_{\rm pec})(1 + z_{\rm TD})(1 + z_{\rm err})\;.
  \label{eq:z_tot}
\end{equation}
In the ``true-centre'' case (no mis-centring) we take the cluster centre to coincide with the centre obtained by the generative model, while in the ``BCG-centre'' case the centre is defined by the Brightest Central Galaxy (BCG). In our mocks where magnitudes are not specified, we use the satellite at the smallest distance from the cluster centre as a proxy for the BCG. We construct the stacked gravitational redshift signal from the mock catalogue by averaging over galaxies and haloes within each projected radial shell. The horizontal error bar assigned to each radial bin corresponds to half of the bin width. 
Finally, we compare our pipeline against the analysis of \citet{2023A&A...669A..29R}, finding consistent uncertainties once similar observational selections are adopted. We choose a true-centre definition, which is closest to the position-average prescription adopted in the above study. We further apply $M_\mathrm{min}=1.5\times10^{14} M_\odot$ and $z_\mathrm{max}=0.5$ to apply the same cuts. We then adopt $C_1=0.7$, $C_2=0.6$ and $\log_{10}(M_\mathrm{cut,0}/M_\odot)=14.7$ to produce a final catalogue consisting of around $3\,000$ clusters and $50\,000$ members, and assume a redshift uncertainty of around $10^{-4}(1+z)$. The above match the cuts in \citet{2023A&A...669A..29R} and the completeness is tuned to reproduce the same number of haloes and member galaxies. Our results for a fixed random seed yield $\alpha_\mathrm{MG}=0.89\pm0.25$, similar in precision to $\alpha_\mathrm{MG}=0.86\pm0.25$ in the aforementioned study. This test shows that the totality of our assumptions, when tuned to mimic those for an observed cluster catalogue, yield reasonable uncertainties.

\begin{table*}
\caption{Summary of the analysis configurations considered in the present study in a self-consistent setting. Mis-specification analyses are presented in Sec. \ref{sec:results}.}
\begin{tabular}{@{}cccccccc@{}}
\toprule
\textbf{variant} & \textbf{$M_\mathrm{min}\;[10^{13}\,M_\odot]$} & \textbf{$z_\mathrm{max}$} & \textbf{galaxy completeness} & \textbf{halo completeness} & \textbf{$\log_{10}[\frac{\sigma_\mathrm{z}}{1+z}]$} & \textbf{$\frac{N_\mathrm{haloes}}{10^6}$} & \textbf{$\frac{N_\mathrm{mem}}{10^7}$} \\ \midrule
0                &                   $3$                   & 2             & 1                            & 1                          & -4                    &        $2.5$          &       $1.5$      \\
1                &               $3$                       &     2          &            1                  &       1                     &   -5                             &    $2.5$                &      $1.5$                 \\
2                &          $3$                            &    2           &       1                       &      1                      &   -6                             &    $2.5$                &    $1.5$                  \\
3                &          $3$                            &    2           &       1                       &      1                      &      -3                          &    $2.5$                &   $1.5$                   \\
4                &        $3$                              &    2           &       1                       &      1                      &        -2                        &    $2.5$                &   $1.5$                   \\\midrule
5                &      $3$                                &    2           &       $C_1=0.9$, $C_2=0.6$   &       1                     &     -4                         &      $2.5$              &      $0.7$               \\
6                &      $3$                                &   2            &       $C_1=0.9$, $C_2=0$  ($\tilde{r}\leq1$), $C_1=0.4$ ($\tilde{r}>1$)              &     1                       &  -4                            &         $2.5$           &       $1$              \\\midrule
7                &     $3$                                 &   2            &          1                    &      $\log_{10}(M_{\rm cut,0}/M_\odot) = 13.8$       &             -4                 &       0.7           &      $0.6$               \\
8               &          $3$                            &     2          &         1                     &       $\log_{10}(M_{\rm cut,0}/M_\odot) = 14.2$                      &    -4                          &    0.3            &    $0.4$ \\\midrule
9                &     $3$                                 &    1           &         1                     &       1                     &           -4                   &   $1.4$                 &     $1$              \\
10                &     $3$                                 &     0.5          &       1                       &     1                       &      -4                        &   0.4                &      $0.3$             \\\midrule
11               &      $5$                                &      2         &      1                        &       1                     &       -4                       &     0.9              &      $0.9$             \\
12               &          $10$                            &       2        &         1                     &         1                   &          -4                    &       0.2            &     $0.4$              \\\midrule
13               &          $10$                            &     2          &         $C_1=0.9$, $C_2=0.6$                     &       $\log_{10}(M_{\rm cut,0}/M_\odot) = 13.8$                     &    -4                          &        0.2          &       $0.2$           \\ 
14               &          $10$                            &     2          &         $C_1=0.9$, $C_2=0.6$                     &       $\log_{10}(M_{\rm cut,0}/M_\odot) = 13.8$                     &    -2                          &        0.2          &        $0.2$           \\\bottomrule
\end{tabular}
\label{tab:variants}
\end{table*}

\section{Results}\label{sec:results}

In our analysis we assess the impact of observational and modelling uncertainties. For the former test, we explore the impact of redshift uncertainties, radial member completeness, maximum redshift and minimum halo mass cut, member galaxy abundance and halo completeness. For the latter test, we explore the impact of mis-specifying the radial density and velocity profiles, the clusters' centres and halo mass function. In each test, we vary a single aspect while keeping all remaining ingredients fixed to the baseline configuration (Table \ref{tab:variants}, variant 0). We set $\alpha_{\rm MG}=1$ throughout our analysis and only show the f(R) and sDGP predictions as reference curves. Our results therefore focus on quantifying phenomenological deviations from the GR prediction.

The baseline scenario is designed to emulate a sample that surpasses the statistical power achievable by current surveys, serving as a benchmark to inform the conceptual planning of next-generation surveys. Our baseline configuration assumes $M_\mathrm{min}=3\times10^{13}M_\odot$ (the minimum mass of the halo catalogue), a redshift range $0<z<2$, perfect redshift completeness for member galaxies, perfect mass–redshift completeness for haloes, a redshift uncertainty of $\sigma_z=10^{-4}(1+z)$, and no mis-specifications in any other component. We further assume a footprint of $18\,000$ deg$^2$, as this will be the area to be covered by the next-generation \texttt{Widefield Spectroscopic Telescope} (WST) cosmology survey \citep{2024arXiv240305398M}, in which efforts are underway for the preparation of a cluster survey \citep{2025arXiv251213221T}. We discuss more upcoming surveys in Section \ref{sec:conclusions}.

\begin{figure}
\centering
\includegraphics[width=0.48\textwidth]{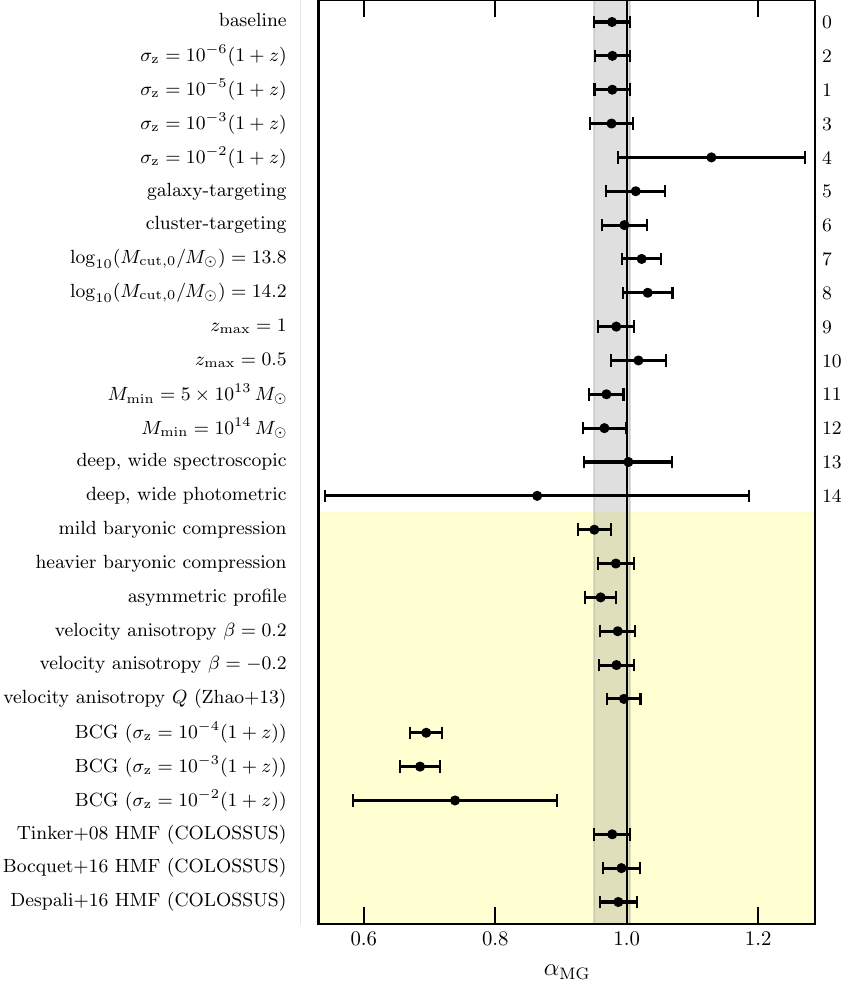}
\vspace{-4mm}
\caption{Summary of the inferred modified gravity parameter for the different variants considered in Sec. \ref{sec:results}, using mocks in GR ($\alpha_{\rm MG}=1$), compared to GR/$\Lambda$CDM (vertical black line). The yellow window highlights modelling uncertainties (model mis-specifications). The baseline case refers to $\sigma_\mathrm{z}=10^{-4}(1+z)$, perfect member and halo completeness for $z_\mathrm{max}=2$ and $M_\mathrm{min}=3\times10^{13}\,M_\odot$. The numbers on the right-hand side vertical axis indicate the variant index each case corresponds to in Table \ref{tab:variants}.}
\label{fig:summary}
\end{figure}

This baseline yields about $ 2.5\times10^{6}$ haloes and $1.5\times10^{7}$ member galaxies. For internal consistency, all stochastic elements (e.g. Gaussian sampling of redshift errors) are generated with a fixed random seed. We provide a summary of the main analysis variants in Table~\ref{tab:variants}, along with the resulting number of haloes and member galaxies. We present a summary of the forecasted constraints from each analysis variant on the modified gravity parameter in Fig. \ref{fig:summary}.

\subsection{Observational uncertainties}\label{sec:unc}

We summarise our results on the impact of observational uncertainties in the white window of Fig. \ref{fig:summary}

\subsubsection{Redshift uncertainty} Here, we investigate the impact of redshift uncertainties of the cluster members on the gravitational redshift signal by comparing variants 0--4 (Table~\ref{tab:variants}), which differ only in the assumed value of $\sigma_z$. For these variants, the number of haloes is $2.5\times10^6$ and the number of members is $1.5\times10^7$. This is an idealised scenario in which all haloes and member galaxies have been identified with perfect completeness. Retaining the original mass threshold of the catalogue, we recover typical uncertainties of around $\sigma_{\bar{\Delta}}=$0.1--1 km $s^{-1}$ on the stacked gravitational redshift signal for spectroscopic redshifts, where $\sigma_\mathrm{z}\leq10^{-3}(1+z)$, across all shells. A sharper trend is observed for photometric surveys, where $\sigma_\mathrm{z}=10^{-2}(1+z)$, in which case the per-shell errors increase to 2--8 km $s^{-1}$. The increase in the gravitational redshift error bars as a function of redshift uncertainty becomes more pronounced in the outermost shell for photometric uncertainties. We fit each radial shell with the empirical scaling $\sigma_{\bar{\Delta}} \propto \sqrt{\sigma_0^2 + B\,\sigma_z^2}\,$ where $\sigma_0$ captures the intrinsic dispersion and $B$ parameterises the propagation of redshift uncertainty, highlighting the transition from the noise floor to the regime dominated by redshift uncertainties. In the regime of large redshift uncertainties, this fit indicates that the gravitational redshift errors grow almost linearly with redshift uncertainty. This behaviour is expected because redshift uncertainties enter the signal linearly. From these fits, we obtain $\sigma_0 \simeq 0.2$--$0.3~\mathrm{km\,s^{-1}}$ across all shells and $B \simeq 10^{4}$--$10^{5}~(\mathrm{km\,s^{-1}})^2$, indicating a sub-km\,s$^{-1}$ noise floor. While spectroscopic uncertainties yield posteriors narrowly centred around the GR expectation, photometric uncertainties broaden the $\alpha_\mathrm{MG}$ posteriors significantly as can be seen in Fig. \ref{fig:summary}. The above findings suggest that the constraining power of gravitational redshifts on the linear modified-gravity parameter improve only marginally for precision better than $\sigma_z \lesssim 10^{-3}(1+z)$. This plateau is naturally explained by the intracluster velocity dispersion, which is of order $450\,\mathrm{km\,s^{-1}}$, corresponding to around $10^{-3}$ in redshift \citep{2001MNRAS.322..901S}. Photometric redshift uncertainties yield $19\%$ precision on $\alpha_\mathrm{MG}$. All other variants yield $\approx 2-4\%$ precision. This indicates that the intrinsic dynamical scatter of the system imposes a threshold: beyond this point, even substantially more precise redshift measurements do not translate into significantly tighter constraints.

\subsubsection{Member completeness} Here, we investigate the impact of the survey's spectroscopic member completeness by comparing variants 0, 5, and 6 in Table \ref{tab:variants} by applying different radial completeness profiles for member galaxies. Variant 5 mimics the completeness of a generic galaxy survey: the spectroscopic completeness starts at $90\%$ near the cluster centre, drops to $50\%$ at $1r_{500}$, and reaches $10\%$ at $4r_{500}$. Variant~6 instead mimics a survey that explicitly targets clusters with the aim of maximising the number of identified members. In this case, the completeness also starts at $90\%$ near the centre, but decreases more slowly with radius: the probability of identifying a galaxy remains $90\%$ out to $1r_{500}$ and only drops to $40\%$ beyond $1r_{500}$. In a realistic survey, such a strategy would typically imply a reduced number of observed clusters, since maintaining this level of completeness is resource-intensive. In this set of tests, the difference between the three configurations in $\sigma_\mathrm{\bar{\Delta}}$ is most pronounced at large radii, because this is the regime in which the galaxy completeness scenarios differ the most. Specifically, the galaxy-targeted completeness performs worst with $\sigma_{\bar{\Delta}} \sim 0.8$ km $s^{-1}$ at $4r_{500}$, the baseline case performs best with $\sigma_{\bar{\Delta}} \sim 0.25$ km $s^{-1}$ at $4r_{500}$, while the cluster-targeted completeness yields $\sigma_{\bar{\Delta}} \sim 0.4$ km $s^{-1}$ at $4r_{500}$. Assuming $100\%$ completeness (variant~0) yields $3\%$ precision on $\alpha_\mathrm{MG}$, while the galaxy-targeted completeness of variant~5 degrades this to $5\%$. The cluster-targeted completeness of variant~6 performs better, yielding $4\%$ precision on $\alpha_\mathrm{MG}$, closer to the idealised $100\%$-completeness case. It is interesting to note that once $\sigma_{\bar{\Delta}}$ becomes less than 0.5 km $s^{-1}$ at $4r_{500}$, there is no additional precision gain on $\alpha_\mathrm{MG}$.

\subsubsection{Halo completeness} Further, we investigate the impact of varying the halo completeness by comparing an idealised, fully complete catalogue (variant~0) with two incomplete cases (variants~7 and~8). Variant~8 is constructed to be more aggressively down-sampled as a function of halo mass and redshift with $\log_{10}(M_{\rm cut,0}/M_\odot)=14.2$ in Eq. \eqref{eq:cluster_completeness} and contains half as many haloes as variant~7, where $\log_{10}(M_{\rm cut,0}/M_\odot)=13.8$. Variant~8 yields $4\%$ precision on $\alpha_\mathrm{MG}$, whereas variant~7 yields $3\%$. The relatively small improvement obtained from a 5-fold increase in the number of haloes is likely driven by the insignificant increase in the number of members. In terms of uncertainty on the observable, this translates into a change of about $0.8\,\mathrm{km\,s^{-1}}$ in $\sigma_{\bar{\Delta}}$ between the least and most conservative completeness assumptions, which as we found above, is close to the sub-km s$^{-1}$ noise floor and thus does not yield any meaningful improvement in the precision of $\alpha_\mathrm{MG}$. The impact of this uncertainty on the $\alpha_\mathrm{MG}$ precision may change under a different parameterisation.

\subsubsection{Maximum redshift} Subsequently, we investigate the impact of the maximum redshift of the cluster catalogue by comparing variants 0, 9, and 10. Imposing a maximum redshift is standard practice to limit the rate of false cluster identifications and to mitigate the impact of possible mis-specification of the cosmological parameters \citep{2023A&A...669A..29R}. As expected, more restrictive cuts lead to looser constraints on $\alpha_\mathrm{MG}$ as they reduce the number of surviving haloes. We find a nonlinear decrease of $\sigma_{\bar{\Delta}}$ with $z_\mathrm{max}$ that tends to saturate for $z_\mathrm{max}>1$. We fit each radial shell with the empirical formula $\sigma_{\bar{\Delta}} \propto \sigma_\infty + b\,z_\mathrm{max}^{-2}$, which encodes an asymptotic error floor $\sigma_\infty$ at high $z_\mathrm{max}$ and the rapid degradation of the precision when the catalogue is truncated at low redshift. From these fits, we obtain $\sigma_\infty \simeq 0.2$ $~\mathrm{km\,s^{-1}}$ and $b \simeq 0.08~\mathrm{km\,s^{-1}}$, with both parameters exhibiting an insignificant increase with radius. This indicates that it is mostly haloes at $z<1$ which contribute to the signal, in line with the domination of the effect by more massive haloes. We find no appreciable difference in the modified-gravity parameter constraints between $z_\mathrm{max}=1$ and $z_\mathrm{max}=2$, both yielding a precision of $3\%$. For a more aggressive cut at $z_\mathrm{max}=0.5$, the precision degrades only to $4\%$.

\subsubsection{Minimum halo mass} Here, we investigate the impact of the lower mass threshold of the cluster catalogue by comparing variants 0, 11, and 12. As above, in a real data setting, this cut would be imposed to reduce the rate of false cluster identifications and exclude the mass range where the richness–mass relation is not reliably calibrated in observations \citep{2023A&A...669A..29R}. We find that $\sigma_{\bar{\Delta}}$ remains close to $0.1$--$0.7$ km s$^{-1}$ across radial distances, but increases linearly with the minimum halo mass cut such that $\sigma_{\bar{\Delta}}\propto M_{\rm min}$. As a result, we find that all three mass thresholds yield very similar constraints on $\alpha_{\rm MG}$, despite the fact that the $M_\mathrm{min}=10^{14}\,M_\odot$ sample contains $10$ times fewer haloes than the $M_\mathrm{min}=3\times10^{13}\,M_\odot$ sample. This behaviour reflects the fact that low-mass haloes generate a much weaker gravitational-redshift signal and therefore contribute negligibly to the overall constraining power even though they are more numerous.

\subsubsection{Member galaxy abundance} In our baseline configuration we assume the HOD corresponding to a galaxy number density of $\bar{n}_\mathrm{g}=0.02\,h^{3}$Mpc$^{-3}$ from \citet[][Table 1]{2005ApJ...633..791Z}, assuming the smooth particle hydrodynamics model. In this part of our analysis we will investigate the impact of varying the number of cluster members by assuming $\bar{n}_\mathrm{g}=0.01\,h^{3}$Mpc$^{-3}$ and $\bar{n}_\mathrm{g}=0.0025\,h^{3}$Mpc$^{-3}$. These scenarios produce $N_\mathrm{mem}=9\times10^6$ and $N_\mathrm{mem}=3\times10^6$ members, respectively. A reduction of order of magnitude $\mathcal{O}(1)$ in the number of members leads to a reduction of $\sigma_{\bar{\Delta}}\simeq 0.3\,\mathrm{km\,s^{-1}}$ in the stacked gravitational redshift uncertainties. We find that the scatter of the stacked signal closely follows a shot-noise scaling, i.e. $\sigma_{\bar{\Delta}} \propto N_{\rm mem}^{-1/2}$ across all projected-radius bins. The precision on the modified gravity parameter degrades from $3\%$ for our baseline HOD and the one for $\bar{n}_\mathrm{g}=0.01\,h^{3}$Mpc$^{-3}$, to $5\%$ for $\bar{n}_\mathrm{g}=0.0025\,h^{3}$Mpc$^{-3}$.

\subsubsection{Photometric vs spectroscopic cluster surveys} Subsequently, we turn to the analysis of variants 13 and 14, which represent deep, wide-field ($18\,000$ deg$^2$) spectroscopic and photometric surveys, respectively, in which the low-mass halo regime has been removed to mimic a selection in a real-data setting. The sample with $\sigma_\mathrm{z}=10^{-4}(1+z)$ yields $7\%$ precision on $\alpha_\mathrm{MG}$ with $\sigma_{\bar{\Delta}}\simeq 0.8-2\,\mathrm{km\,s^{-1}}$, whereas the photometric-like sample with $\sigma_\mathrm{z}=10^{-2}(1+z)$ yields $56\%$ precision on $\alpha_\mathrm{MG}$ with $\sigma_{\bar{\Delta}}\simeq 6$--$46\,\mathrm{km\,s^{-1}}$. This indicates that, for halo abundances of order $10^5$, spectroscopic redshifts are required to drive the statistical uncertainties on the stacked signal down to the few-km\,s$^{-1}$ regime; deep photometry alone is not sufficient. We further perform a comparison between a deep, wide-field photometric mock survey ($N_\mathrm{haloes}=180\,000$, $z_\mathrm{max}=2$) and a narrower, shallower ($9\,000$ deg$^2$, $N_\mathrm{haloes}=60\,000$, $z_\mathrm{max}=0.5$) spectroscopic one. While in the former case we obtain $56\%$ precision on $\alpha_\mathrm{MG}$, in the latter case we find $10\%$ precision. 
This indicates that, for cluster gravitational-redshift measurements, a narrower and shallower spectroscopic survey is preferable to a deep, wide-field photometric survey.

\begin{figure}
\centering
\includegraphics[width=0.45\textwidth]{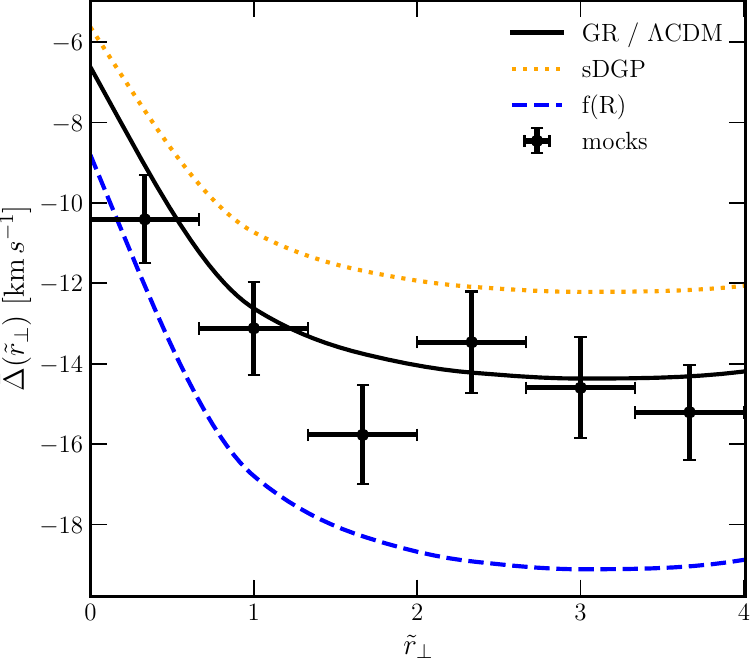}
\caption{Forecasted uncertainty on the stacked gravitational redshift profile for a magnitude-limited spectroscopic cluster catalogue with $m_{\rm lim}=23$ mag, obtained from a galaxy survey out to $z_\mathrm{max}=1$ assuming a member completeness of $C_1=0.9$, $C_2=0.6$, $M_\mathrm{min}=10^{14}$ M$_\odot$ and $\log_{10}(M_{\rm cut,0}/M_\odot)=13.8$. This represents a relatively resource-efficient setting for Stage-V surveys which delivers $4\%$ precision on $\alpha_{\rm MG}$ in the absence of any mis-specifications. The mock data are generated assuming GR.}
\label{fig_9b}
\end{figure}

Finally, we explore which survey configuration could be most informative for inferences of modified gravity with cluster gravitational redshifts, yet resource-efficient, in the absence of mis-specifications. We consider a spectroscopic survey with $\sigma_\mathrm{z}=10^{-4}(1+z)$ extending to $z_\mathrm{max}=0.5$ on which a lower mass limit of $10^{14}$ M$_\odot$ is imposed to mitigate any false cluster identification effects which would be present in a real catalogue. Applying the halo completeness configuration of variant 13, this setting produces around $10^5$ clusters with $3\times10^5$ member galaxies. Under perfect knowledge of the clusters' centres, we obtain $20\%$ and $14\%$ precision on $\alpha_{\rm MG}$ for a galaxy- and cluster-targeting survey, respectively (member completeness of variants 5, 6, respectively). Extending such a galaxy survey to $z_\mathrm{max}=1$ improves the $\alpha_{\rm MG}$ precision from $14\%$ to $10\%$, whereas for a cluster survey we obtain $8\%$ precision on $\alpha_{\rm MG}$. We therefore propose a spectroscopic galaxy survey out to $z_\mathrm{max}=1$ from which a spectroscopic cluster catalogue is built as the most resource-efficient configuration for future real-data applications among the survey specifications considered. Notice that the impact of maximum redshift and member completeness becomes more prominent as the minimum mass cut increases. This is because all haloes in the catalogue contribute significantly to the signal. 

\subsection{Modelling uncertainties}\label{sec:mis}

\begin{figure*}
\centering
\includegraphics[width=\textwidth]{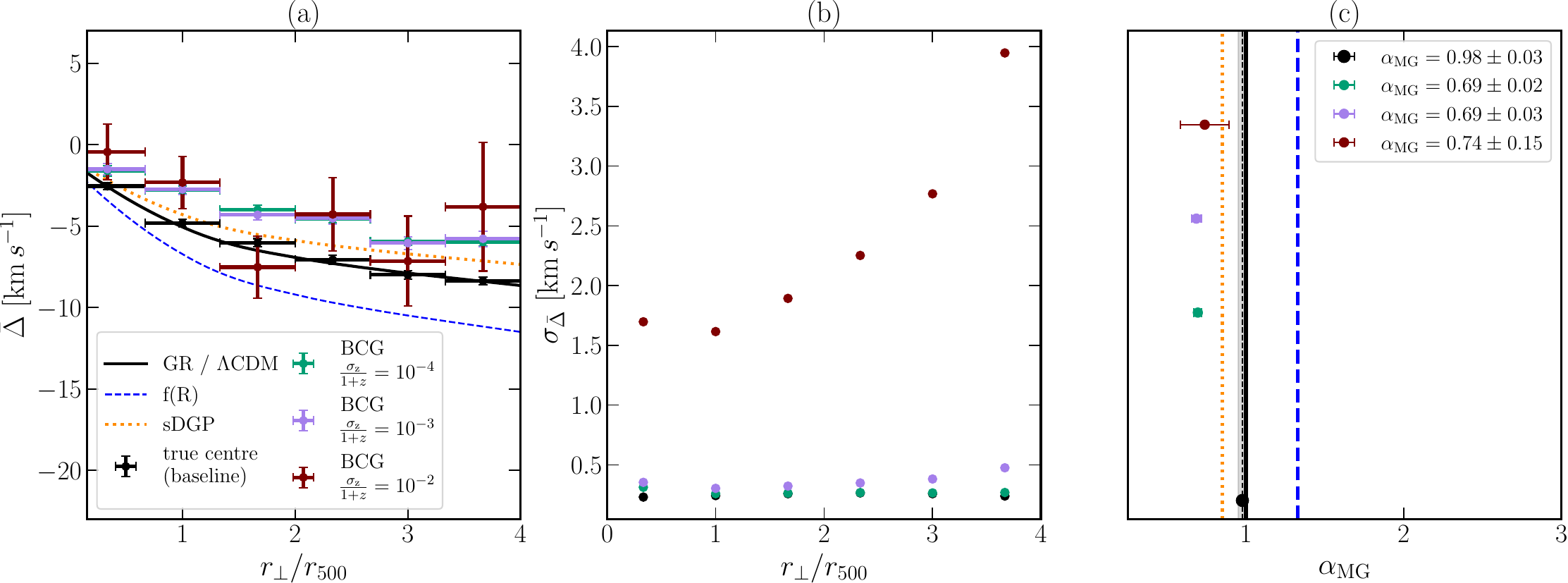}
\vspace{-4mm}
\caption{(a) Gravitational redshift signal and (b) error on the signal as a function of radial distance from the clusters' centres' and (c) corresponding modified gravity parameter for different redshift uncertainties on the BCG, assuming spectroscopic uncertainty for the member galaxies.}
\label{fig_9}
\end{figure*}

We summarise our results on the impact of observational uncertainties in the white region of Fig. \ref{fig:summary}, and provide more details where appropriate in Figs. \ref{fig_9b} and \ref{fig_9}.

\subsubsection{Radial profile} We generate mock catalogues in which the true halo profile deviates from NFW according to the two symmetric profiles based on Eq.~\eqref{eq:nfw_bar}, while the analysis model incorrectly assumes the NFW form of Eq.~\eqref{eq:nfw_rho}. The three cases do not differ significantly in their $\sigma_{\bar{\Delta}}$ error bars, yielding $3\%$ precision on $\alpha_{\rm MG}$. The case of a sample with partially asymmetric profiles modelled with a neighbouring, less massive halo also produces the same precision with no appreciable bias with respect to the baseline configuration, at least given the linear modified gravity parameterisation. In the regime of a massive spectroscopic cluster survey, as emulated by our baseline configuration, such mis-specifications of the radial density profile do not produce clearly visible systematic offsets between the measured gravitational redshift signal and the (wrong) NFW-based prediction. Our results therefore indicate that even when observational uncertainties reach the regime $\sigma_{\bar{\Delta}}\simeq \mathcal{O}(1)\,\mathrm{km\,s^{-1}}$, mis-modelling of the halo density profile cannot significantly mimic linear departures from GR. We note that \citet{2017MNRAS.468.1981C} found that an NFW-only model applied to a sample with asymmetric density profiles can overpredict the gravitational-redshift signal. This motivates extending the present toy asymmetry test to mock catalogues drawn from full cosmological environments, where substructure, neighbouring haloes, and the cosmic web are modelled self-consistently.

\subsubsection{Velocity dispersion anisotropy} We then investigate the impact of introducing anisotropy in the mock velocity dispersion profiles while ignoring it in the model. We find a negligible difference in $\sigma_{\bar{\Delta}}$ among the $\beta=0.2$, $\beta=-0.2$ and $\beta=0$ (isotropic) cases. As a result, the precision in the modified gravity parameter does not change significantly between these observationally-motivated scenarios and remains at $3\%$ and consistent with GR. We further investigate the impact of keeping $\beta=0$ and breaking the assumption of isotropic galaxy orbits by varying the baseline $Q=3/2$ as discussed in Sec. \ref{sec:model}. We obtain $\alpha_{\rm MG}=1.00\pm0.03$, consistent with both the anisotropy and lack thereof scenarios. Further, we test the scaling of $\sigma_{\bar{\Delta}}$ with the magnitude of the velocity dispersion by amplifying the baseline velocity dispersion $\sigma$. We find a $\sigma_{\bar{\Delta}}\propto\sigma$ scaling, indicating a linear increase of the error bars in velocity models with greater predicted velocity dispersion. This is because velocities are added linearly to the signal for the dominant gravitational redshift contribution.

\subsubsection{Cluster centre mis-specification} For the next part of our analysis, we examine the impact of mis-specifying the cluster centre. Our baseline configuration assumes a perfectly known centre. In Fig.~\ref{fig_9}, we show how our results change when the centre is instead taken to be the galaxy with the smallest projected distance from the halo centre (the mock BCG), whose redshift is assigned different levels of uncertainty. If cluster centres are chosen to coincide with the location of member galaxies, our setting mimics the least-offset scenario per cluster. When the BCG is used as a purely geometric proxy for the centre, even with a redshift known within $\sigma_z=10^{-4}(1+z)$, the derived modified gravity parameter is systematically offset from the ground truth. As the BCG redshift uncertainty is increased to $10^{-2}(1+z)$, the error bars on $\bar{\Delta}_\mathrm{gz}$ grow progressively, as shown in Fig.~\ref{fig_9}(b), especially in the cluster outskirts. The posterior on the modified-gravity parameter $\alpha_\mathrm{MG}$ broadens correspondingly as shown in Fig.~\ref{fig_9}(c). For $\sigma_z=10^{-2}(1+z)$, the centring noise dominates: the radial errors inflate to several km\,s$^{-1}$ and the posterior on $\alpha_\mathrm{MG}$ becomes wider. These results indicate that, in a high-precision spectroscopic setting, geometric mis-centring on the BCG cannot be ignored. The systematic shift towards more positive values of $\bar{\Delta}_\mathrm{gz}$ is also reported by \citet{2023A&A...669A..29R}. This arises because, if we define the reference potential at $r_{\rm cen}>0$ instead of at the true centre, the measured signal becomes $\Delta_{\rm cen}(\tilde{r})=\phi(\tilde{r}_{\rm cen})-\phi(\tilde{r})=\Delta_{\rm true}(\tilde{r})-\Delta_{\rm true}(\tilde{r}_{\rm cen})=\Delta_{\rm true}(\tilde{r})+|\Delta_{\rm true}(\tilde{r}_{\rm cen})|$, i.e. a strictly positive, radius-independent offset is added to the true profile. In Sec. \ref{sec:conclusions} we propose mitigation strategies.

\subsubsection{Halo mass function mis-specification} For the final part of this analysis, we examine the impact of mis-specifying the HMF. In our baseline configuration we adopt the \citet{Tinker08} HMF evaluated at the true redshift of each cluster with \codefont{CosmoBolognaLib}\footnote{\href{https://gitlab.com/federicomarulli/CosmoBolognaLib}{gitlab.com/federicomarulli/CosmoBolognaLib}} (\codefont{CBL})  \citep{2016A&C....14...35M} for the critical $M_{500}$ mass definition. Here, we compare results obtained with different HMF prescriptions for the critical $M_{500}$ mass definition with \codefont{COLOSSUS} \citep{2018ApJS..239...35D}. In particular, we contrast the Tinker model with the \citet{2016MNRAS.456.2361B} and \citet{2016MNRAS.456.2486D} variants. The mis-specification consists both of physical differences between the models, but also numerical differences between \codefont{COLOSSUS} and \codefont{CBL}. To account for the latter, in Fig. \ref{fig:summary} we shift the \codefont{COLOSSUS} mean prediction for Tinker to match the mean of the \codefont{CBL} prediction. Apart from numerical differences between \codefont{COLOSSUS} and \codefont{CBL} which impact the signal negligibly, mis-specifying the HMF does not yield any appreciable discrepancy from the fiducial GR scenario. Using the sample's HMF instead of the \codefont{CBL} one produces identical model predictions.

In Sec. \ref{sec:unc} we saw that when a more realistic lower mass threshold is imposed, effects which had a smaller impact in our baseline configuration become important. We therefore repeat the above investigations in the setting of the survey presented in Fig. \ref{fig_9b}. We find that the above mis-specifications, with the exception of the cluster centre, still do not mimic any significant departures from GR. This finding validates the assumption of minimal impact of mis-specifications on the gravitational redshift signal in a real-data setting which was concluded through experimentation with data cuts by \citet{2023A&A...669A..29R}. Further, it highlights that, looking ahead to Stage-IV and Stage-V survey exploitation, the primary modelling focus should be on centre mis-specification. We summarise the findings of the above investigations in Fig. \ref{fig:summary}.

\subsection{Flux-limited survey forecast}\label{sec:SB}

In the previous sections, we found that a $z<1$ spectroscopic survey targeting galaxies emerges as a cost-effective configuration. Here, we assume a limiting redshift of $z<1$, $\sigma_z=10^{-4}(1+z)$, a spectroscopic member completeness mimicking that of a galaxy survey ($C_1=0.9$, $C_2=0.6$), a halo completeness with $\log_{10}(M_{\rm cut,0}/M_\odot)=13.8$, a minimum mass cut of $M_{\rm min}=10^{14}$ $M_\odot$, and an apparent magnitude limit of $m_{\rm lim}=23$ mag. The incorporation of the latter effect requires us to account for surface brightness modulation (relativistic beaming), as discussed in App. \ref{app:SB}. In this setting, we end up with around $10^5$ haloes and $10^6$ members, assuming an HOD that reproduces the observed galaxy number density for the corresponding magnitude limit in \citet{2007ApJ...667..760Z}. Incorporating the effect introduces a preferential selection of galaxies moving toward us that translate into a mean bias of around $\langle z_{\rm pec}\rangle = -10^{-5}\pm10^{-6}$ across galaxies. We present our forecast assuming no mis-centring in Fig. \ref{fig_9b}. We find $4\%$ constraints on $\alpha_\mathrm{MG}$, consistent with the fiducial GR scenario. We vary the amplitude of the effect by assuming $m_{\rm lim}=22$ mag following \citet{2023A&A...669A..29R}. We find $5\%$ constraints on $\alpha_\mathrm{MG}$, consistent with the fiducial GR scenario. For the final part of our analysis, we create a setup like the one assumed above for $m_\mathrm{lim}=23$ mag, adding mis-centring from a spectroscopically observed BCG with $\sigma_\mathrm{z}=10^{-4}(1+z)$, asymmetry and mild baryonic compression in the radial density profiles, as well as velocity anisotropy with $\beta=0.2$. We find $\alpha_\mathrm{MG}=0.79\pm0.04$. Therefore, we find that there is no compensating effect among the ones we considered that could mask the impact of mis-centring.

\section{Conclusions}\label{sec:conclusions}

In the present study we developed an end-to-end pipeline that takes a halo catalogue as input, populates it with galaxies according to an HOD and a prescribed radial density profile, predicts the corresponding gravitational redshift signal in stacked cluster samples, and infers deviations from GR via a simple modified-gravity parametrisation that rescales the gravitational constant in GR. The resulting framework is a tool that can be tuned to survey specifications and used for fast, survey-specific forecasts, like the one presented in \citet{2025arXiv251213221T}. 

The data vector we considered consists of the halo properties, $(M_{500}, r_{500}, c_{500})$ and redshifts, and the member properties, $(\tilde{r}, z, z_\mathrm{pec})$, along with their sky coordinates and redshift uncertainties. Our model further assumes a member and halo completeness, as well as a halo occupation distribution. In our analysis, we showed that, given these inputs, we are able to model the gravitational redshift signal accurately in the limit of infinitesimal uncertainties. We used the validated pipeline to explore the impact of a range of observational and modelling systematic effects. In particular, we investigated the effects of redshift uncertainties, member-galaxy and halo completeness, maximum redshift and minimum halo-mass cuts, member galaxy abundance, and the mis-specifications of the radial density profile, cluster centres, velocity dispersion anisotropy and halo mass function. Our main findings can be summarised as follows.
\begin{itemize}
    \item There exists an effective floor, set by the intracluster velocity dispersion ($\sigma_v \sim 450\,\mathrm{km\,s^{-1}}$), below which improving redshift precision does not translate into tighter constraints on $\alpha_\mathrm{MG}$.
    \item Surveys that maximise spectroscopic completeness for cluster members can push modified-gravity constraints beyond what is achievable with generic galaxy surveys, yet the precision gain is $\sim 1\%$ on the modified gravity parameter.
    \item Redshift and halo-mass cuts designed to reduce the rate of false cluster identification do not substantially weaken the constraining power, as they primarily remove low-mass haloes that contribute little to the signal.
    \item Consistently with this, realistic halo completeness as a function of mass and redshift has only a modest impact on the $\alpha_\mathrm{MG}$ constraints.
    \item Mis-specifying the radial density profile by neglecting baryonic contributions does not bias the predicted signal. This conclusion is limited to the profile variations tested here, which preserve $M_{500}$ and mainly affect the inner density structure. These perturbations do not project strongly onto the linear amplitude parameter $\alpha_{\rm MG}$, although more general profile mis-modelling in self-consistent beyond-GR mocks could produce stronger degeneracies. This investigation should be extended to full $N$-body simulations.
    \item Mis-specifying the cluster centres by using the BCG as a purely geometric proxy introduces a significant systematic positive shift in the signal, mimicking deviations from GR. 
    \item Narrow and shallow spectroscopic samples are preferable against wide and deep photometric cluster samples for precise $\alpha_\mathrm{MG}$ inference.
    \item A $z<1$ spectroscopic survey targeting galaxies emerges as the most cost-effective configuration for constraining $\alpha_\mathrm{MG}$ with gravitational redshifts.
    \item The only mis-specification which can mimic signatures of modified gravity is that of cluster mis-centering up to $15\sigma$ toward $f(R)$ gravity.
    \item The uncertainties on the stacked gravitational redshift signal scale with the velocity dispersion and number of tracers per shell roughly as $\sigma_{\bar{\Delta}}(\tilde{r}_{\perp, \rm i})\propto \sigma(\tilde{r}_{\perp, \rm i})/\sqrt{N_\mathrm{mem,i}}$ in a radial distance bin $i$.
\end{itemize}
From a modelling perspective, we showed that velocity dispersion anisotropies can be accounted for in the model by modifying Eq. \eqref{eq:fTD}, even when the anisotropy parameter becomes a function of radial distance as shown in \citet[][Fig. 5]{2025ApJ...987...70A} and \citet{2026A&A...707A.153B, 2026arXiv260215934M}. 

As statistical error bars shrink, careful modelling will be essential to avoid spurious deviations from GR. Our results highlight mis-centring as a potentially significant systematic effect in gravitational-redshift analyses. The centring prescription adopted here should not be interpreted as a physical model of BCG formation or luminosity selection. In our mocks, where the HOD yields a limited number of members per halo, we instead adopt the member galaxy with the smallest offset from the halo centre as a minimal centring proxy. The sensitivity of the inferred signal to the centring prescription motivates future calibration with cluster-galaxy mocks based on hydrodynamical simulations or on semi-analytic/empirical galaxy-population models applied to $N$-body simulations. Therefore, efforts should be invested into constructing an accurate proxy for cluster centres.

\citet{2023A&A...669A..29R} found that the average of member positions can produce a more reliable estimate of a cluster's minimum of the gravitational potential using an empirical richness cut at $\lambda=3$. Below this value, the performance of the estimator is similar to considering the BCG as the cluster centre, whereas for $\lambda>5$ the statistics become too low to obtain accurate measurements. They attributed this effect to the fact that the BCG could be misidentified due to the surface brightness modulation effect. \citet{2025arXiv251214636A} reported that a luminosity-weighted average of the cluster member positions yields a less accurate centre proxy than the BCG. Given these findings, and the fact that we find the centre mis-specification to be the only significant systematic effect for next-generation inferences of modified gravity from gravitational redshifts, it is paramount that future efforts be invested into (i) exploring different averaging prescriptions, (ii) determining a cluster's centre from X-ray and optical observations jointly \citep{2023A&A...671A..57S}, (iii) obtaining additional information with integral field spectroscopy, or from a statistical perspective, (iv) adopting a prior on the centre, rather than fixing it. In the latter direction, the offset can be approximated as a Gaussian shift of the halo relative to the assumed centre \citep{2007arXiv0709.1159J,2015MNRAS.452.3529V,2019MNRAS.484.1598B}. Among the strategies considered above, adopting a free parameter for the cluster centre to be marginalised over emerges as the most straightforward option statistically. In this case, degeneracies between centring and gravity modifications can be broken by leveraging measurements across multiple radial bins. Varying the fiducial value of $\alpha_{\rm MG}$ does not alter our conclusions regarding the relative impact of the observational uncertainties considered here. Among modelling uncertainties, mis-centring remains the dominant source of bias, with an amplitude that scales with the fiducial value of $\alpha_{\rm MG}$. This check should, however, be repeated with fully self-consistent beyond-GR mocks, since modified gravity may also alter the sample properties, as discussed in Section~\ref{sec:model}.

Achieving percent-level precision on the modified gravity parameter is demanding, requiring both very large cluster samples and substantial spectroscopic sampling within each system. This naturally points to a combination of wide-area photometric surveys such as Euclid or Rubin for cluster identification, together with highly multiplexed spectroscopic follow-up capable of targeting $\mathcal{O}(10^5)$ clusters. In this context, Stage-V facilities such as the WST and \texttt{The Spectroscopic Stage-5 Experiment} (Spec-S5) \citep{2025arXiv250307923B} are particularly compelling. Spec-S5, with its pair of 6m telescopes delivering around 13\,000 simultaneous spectra at high spatial density, represents a major advance over current capabilities and, for the first time, enables cluster spectroscopy at the scale required for precision tests of gravity, even though multiple passes may still be needed in the densest regions. WST, combining large aperture and high multiplex is the best long-term prospect, especially if it has close fibre spacing (around $7\,\arcsec$). Together, these facilities make a clear path toward exploiting gravitational redshifts as a competitive probe of modified gravity. In the meantime, combining existing spectroscopic datasets with current and forthcoming photometric surveys can already extend the reach of existing measurements and provide important constraints.

Our results also underscore the need for high-resolution modified-gravity simulations in large volumes: departures from GR affect not only the gravitational redshift signal but also the halo population itself, which in this work is still drawn from a GR-based catalogue. In this context, it is crucial to develop halo catalogues in modified gravity that are populated self-consistently with member galaxies \citep{2015JCAP...12..049S,2019JCAP...03..020S,10.1093/mnras/stac1528,2025MNRAS.538.1415S,2026MNRAS.545f2086V}. While significant progress has been made in fast modelling of structure formation \citep[e.g.][]{2018MNRAS.481.2813G,2020MNRAS.497.1885H,2021JCAP...09..024B,2022JCAP...05..018R,2025A&A...695A.170B}, these efforts now need to be pushed into the regime of gravity models that remain viable after GW170817, avoid excessive tuning and ghost instabilities, remain unscreened on the relevant cluster scales and densities, and preferably admit a healthy self-accelerating branch.

Future work should further focus on propagating halo-mass uncertainties from weak-lensing calibration into the predicted signal, and on configuring mock catalogues of the type presented here to match the specific selection functions and systematic effect budgets of individual surveys. Finally, our analysis indicates that simple linear rescalings of the gravitational constant are not especially informative as a parameterisation of modified gravity in this context: even under optimistic assumptions, the best constraints we obtain are at the $3\%$ level. This motivates exploring more modified gravity parameterisations \citep[e.g.][]{2013PhRvD..88h3513N,18,2022PDU....3701069J}, with the aim of making fuller use of the constraining power of upcoming surveys. 

\section*{Acknowledgements}\label{acknowledgements}
This work was supported by STFC through Imperial College Astrophysics Consolidated Grant ST/W000989/1. ET acknowledges support from the FoNS Researcher Mobility Grant 2025. 

\section*{Data availability}

Data products underlying this article can be made available upon reasonable request to the corresponding author. 



\bibliographystyle{mnras}
\bibliography{main} 

@ARTICLE{2021CQGra..38o3001D,
       author = {{Di Valentino}, Eleonora and {Mena}, Olga and {Pan}, Supriya and {Visinelli}, Luca and {Yang}, Weiqiang and {Melchiorri}, Alessandro and {Mota}, David F. and {Riess}, Adam G. and {Silk}, Joseph},
        title = "{In the realm of the Hubble tension-a review of solutions}",
      journal = {Classical and Quantum Gravity},
         year = 2021,
        month = jul,
       volume = {38},
       number = {15},
          eid = {153001},
        pages = {153001},
          doi = {10.1088/1361-6382/ac086d},
archivePrefix = {arXiv},
       eprint = {2103.01183},
 primaryClass = {astro-ph.CO},
       adsurl = {https://ui.adsabs.harvard.edu/abs/2021CQGra..38o3001D}
}

@ARTICLE{2025arXiv250307923B,
       author = {{Besuner}, Robert and {Dey}, Arjun and {Drlica-Wagner}, Alex and {Ebina}, Haruki and {Fernandez Moroni}, Guillermo and {Ferraro}, Simone and {Forero-Romero}, Jaime and {Honscheid}, Klaus and {Jelinsky}, Pat and {Lang}, Dustin and {Levi}, Michael and {Martini}, Paul and {Myers}, Adam and {Palanque-Delabrouille}, Nathalie and {Panda}, Swayamtrupta and {Poppett}, Claire and {Sailer}, Noah and {Schlegel}, David and {Shafieloo}, Arman and {Silber}, Joseph and {White}, Martin and {Abbott}, Timothy and {Allen}, Lori and {Avila}, Santiago and {Avil{\'e}s}, Roberto and {Bailey}, Stephen and {Bault}, Abby and {Bouri}, Mohamed and {Boutsia}, Konstantina and {Burtin}, Eienne and {Chierchie}, Fernando and {Coulton}, William and {Dawson}, Kyle and {Dey}, Biprateep and {Dor{\'e}}, Olivier and {Dunlop}, Patrick and {Eisenstein}, Daniel and {Emanuele}, Castorina and {Escoffier}, Stephanie and {Estrada}, Juan and {Fagrelius}, Parker and {Fanning}, Kevin and {Fanning}, Timothy and {Font-Ribera}, Andreu and {Frieman}, Joshua and {Galal}, Malak and {Gluscevic}, Vera and {Gontcho}, Satya Gontcho A and {Green}, Daniel and {Gutierrez}, Gaston and {Guy}, Julien and {Hashemi}, Kevan and {Heathcote}, Steve and {Holland}, Steve and {Hou}, Jiamin and {Huterer}, Dragan and {Irigoyen Gimenez}, Blas and {Ivanov}, Mikhail and {Joyce}, Richard and {Jullo}, Eric and {Juneau}, Stephanie and {Juramy}, Claire and {Karcher}, Armin and {Kent}, Stephen and {Kirkby}, David and {Kneib}, Jean-Paul and {Krause}, Elisabeth and {Krolewski}, Alex and {Lahav}, Ofer and {Lapi}, Agustin and {Leauthaud}, Alexie and {Lewandowski}, Matthew and {Li}, Ting and {Lin}, Kenneth and {Loverde}, Marilena and {MacBride}, Sean and {Magneville}, Christophe and {Marshall}, Jennifer and {McDonald}, Patrick and {Miller}, Timothy and {Moustakas}, John and {M{\"u}nchmeyer}, Moritz and {Najita}, Joan and {Newman}, Jeff and {Percival}, Will and {Philcox}, Oliver and {Pires}, Priscila and {Raichoor}, Anand and {Roach}, Brandon and {Rockosi}, Constance and {Rombach}, Maxime and {Ross}, Ashley and {Sanchez}, Eusebio and {Schmidt}, Luke and {Schubnell}, Michael and {Sebok}, Rebekah and {Seljak}, Uros and {Silverstein}, Eva and {Slepian}, Zachay and {Stone}, Chris and {Stupak}, Robert and {Tarl{\'e}}, Gregory and {Li}, Ting and {Tyas}, Luke and {Vargas-Maga{\~n}a}, Mariana and {Walker}, Alistair and {Wenner}, Nicholas and {Y{\`e}che}, Christophe and {Zhang}, Yuanyuan and {Zhou}, Rongpu},
        title = "{The Spectroscopic Stage-5 Experiment}",
      journal = {arXiv e-prints},
         year = 2025,
        month = mar,
          eid = {arXiv:2503.07923},
        pages = {arXiv:2503.07923},
          doi = {10.48550/arXiv.2503.07923},
archivePrefix = {arXiv},
       eprint = {2503.07923},
 primaryClass = {astro-ph.CO},
       adsurl = {https://ui.adsabs.harvard.edu/abs/2025arXiv250307923B}
}

@ARTICLE{2017PhRvL.119y1304E,
       author = {{Ezquiaga}, Jose Mar{\'\i}a and {Zumalac{\'a}rregui}, Miguel},
        title = "{Dark Energy After GW170817: Dead Ends and the Road Ahead}",
      journal = {\prl},
         year = 2017,
        month = dec,
       volume = {119},
       number = {25},
          eid = {251304},
        pages = {251304},
          doi = {10.1103/PhysRevLett.119.251304},
archivePrefix = {arXiv},
       eprint = {1710.05901},
 primaryClass = {astro-ph.CO},
       adsurl = {https://ui.adsabs.harvard.edu/abs/2017PhRvL.119y1304E}
}

@ARTICLE{2020MNRAS.497.1885H,
       author = {{Hassani}, Farbod and {Lombriser}, Lucas},
        title = "{N-body simulations for parametrized modified gravity}",
      journal = {\mnras},
         year = 2020,
        month = sep,
       volume = {497},
       number = {2},
        pages = {1885-1894},
          doi = {10.1093/mnras/staa2083},
archivePrefix = {arXiv},
       eprint = {2003.05927},
 primaryClass = {astro-ph.CO},
       adsurl = {https://ui.adsabs.harvard.edu/abs/2020MNRAS.497.1885H}
}

@ARTICLE{2025A&A...695A.170B,
       author = {{Breton}, Michel-Andr{\`e}s},
        title = "{PySCo: A fast particle-mesh N-body code for modified gravity simulations in Python}",
      journal = {\aap},
         year = 2025,
        month = mar,
       volume = {695},
          eid = {A170},
        pages = {A170},
          doi = {10.1051/0004-6361/202452770},
archivePrefix = {arXiv},
       eprint = {2410.20501},
 primaryClass = {astro-ph.CO},
       adsurl = {https://ui.adsabs.harvard.edu/abs/2025A&A...695A.170B}
}

@ARTICLE{2024MNRAS.527.7242S,
       author = {{S{\'a}ez-Casares}, I{\~n}igo and {Rasera}, Yann and {Li}, Baojiu},
        title = "{The e-MANTIS emulator: fast predictions of the non-linear matter power spectrum in f(R)CDM cosmology}",
      journal = {\mnras},
         year = 2024,
        month = jan,
       volume = {527},
       number = {3},
        pages = {7242-7262},
          doi = {10.1093/mnras/stad3343},
archivePrefix = {arXiv},
       eprint = {2303.08899},
 primaryClass = {astro-ph.CO},
       adsurl = {https://ui.adsabs.harvard.edu/abs/2024MNRAS.527.7242S}
}

@ARTICLE{2018PhRvD..97f1501L,
       author = {{Langlois}, David and {Saito}, Ryo and {Yamauchi}, Daisuke and {Noui}, Karim},
        title = "{Scalar-tensor theories and modified gravity in the wake of GW170817}",
      journal = {\prd},
         year = 2018,
        month = mar,
       volume = {97},
       number = {6},
          eid = {061501},
        pages = {061501},
          doi = {10.1103/PhysRevD.97.061501},
archivePrefix = {arXiv},
       eprint = {1711.07403},
 primaryClass = {gr-qc},
       adsurl = {https://ui.adsabs.harvard.edu/abs/2018PhRvD..97f1501L}
}

@ARTICLE{2025JCAP...09..047M,
       author = {{Mironov}, S. and {Sharov}, M. and {Volkova}, V.},
        title = "{Time-dependent, spherically symmetric background in Kaluza-Klein compactified Horndeski theory and the speed of gravity waves}",
      journal = {\jcap},
         year = 2025,
        month = sep,
       volume = {2025},
       number = {9},
          eid = {047},
        pages = {047},
          doi = {10.1088/1475-7516/2025/09/047},
archivePrefix = {arXiv},
       eprint = {2408.06329},
 primaryClass = {gr-qc},
       adsurl = {https://ui.adsabs.harvard.edu/abs/2025JCAP...09..047M}
}

@ARTICLE{2025PDU....4801906G,
       author = {{Giar{\`e}}, William and {Mahassen}, Tariq and {Valentino}, Eleonora Di and {Pan}, Supriya},
        title = "{An overview of what current data can (and cannot yet) say about evolving dark energy}",
      journal = {Physics of the Dark Universe},
         year = 2025,
        month = may,
       volume = {48},
          eid = {101906},
        pages = {101906},
          doi = {10.1016/j.dark.2025.101906},
archivePrefix = {arXiv},
       eprint = {2502.10264},
 primaryClass = {astro-ph.CO},
       adsurl = {https://ui.adsabs.harvard.edu/abs/2025PDU....4801906G}
}

@ARTICLE{2024MNRAS.532.3972L,
       author = {{Lyall}, Stuart and {Blake}, Chris and {Turner}, Ryan J.},
        title = "{Constraining modified gravity scenarios with the 6dFGS and SDSS galaxy peculiar velocity data sets}",
      journal = {\mnras},
         year = 2024,
        month = aug,
       volume = {532},
       number = {4},
        pages = {3972-3984},
          doi = {10.1093/mnras/stae1718},
archivePrefix = {arXiv},
       eprint = {2407.18684},
 primaryClass = {astro-ph.CO},
       adsurl = {https://ui.adsabs.harvard.edu/abs/2024MNRAS.532.3972L}
}

@ARTICLE{2024MNRAS.534..349L,
       author = {{Landim}, Ricardo G. and {Desmond}, Harry and {Koyama}, Kazuya and {Penny}, Samantha},
        title = "{Testing screened modified gravity with SDSS-IV-MaNGA}",
      journal = {\mnras},
         year = 2024,
        month = oct,
       volume = {534},
       number = {1},
        pages = {349-360},
          doi = {10.1093/mnras/stae2096},
archivePrefix = {arXiv},
       eprint = {2407.08825},
 primaryClass = {astro-ph.CO},
       adsurl = {https://ui.adsabs.harvard.edu/abs/2024MNRAS.534..349L}
}

@ARTICLE{2025arXiv251205819V,
       author = {{Vogt}, S.~M.~L. and {Bocquet}, S. and {Davies}, C.~T. and {Mohr}, J.~J. and {Schmidt}, F. and {Ruan}, C.-Z. and {Li}, B. and {Hern{\"a}ndez-Aguayo}, C. and {Grandis}, S. and {Bleem}, L.~E. and {Klein}, M. and {Aguena}, M. and {Allam}, S. and {Andrade-Oliveira}, F. and {Bacon}, D. and {Brooks}, D. and {Camilleri}, R. and {Carnero Rosell}, A. and {Carretero}, J. and {Costanzi}, M. and {da Costa}, L.~N. and {da Silva Pereira}, M.~E. and {De Vicente}, J. and {Doel}, P. and {Garc{\"\i}a-Bellido}, J. and {Giles}, P. and {Gruen}, D. and {Gutierrez}, G. and {Hinton}, S.~R. and {Hollowood}, D.~L. and {James}, D.~J. and {Kuehn}, K. and {Lee}, S. and {Marshall}, J.~L. and {Mena-Fern{\"a}ndez}, J. and {Menanteau}, F. and {Miquel}, R. and {Myles}, J. and {Plazas Malag{\"o}n}, A.~A. and {Porredon}, A. and {Prat}, J. and {Reichardt}, C.~L. and {Romer}, A.~K. and {Sanchez}, E. and {Sevilla-Noarbe}, I. and {Smith}, M. and {Soares-Santos}, M. and {Suchyta}, E. and {Swanson}, M.~E.~C. and {To}, C. and {Vikram}, V. and {Weaverdyck}, N.},
        title = "{Constraints on nDGP gravity from SPT galaxy clusters with DES and HST weak-lensing mass calibration and from Planck PR4 CMB anisotropies}",
      journal = {arXiv e-prints},
         year = 2025,
        month = dec,
          eid = {arXiv:2512.05819},
        pages = {arXiv:2512.05819},
          doi = {10.48550/arXiv.2512.05819},
archivePrefix = {arXiv},
       eprint = {2512.05819},
 primaryClass = {astro-ph.CO},
       adsurl = {https://ui.adsabs.harvard.edu/abs/2025arXiv251205819V}
}

@ARTICLE{2011Natur.477..567W,
       author = {{Wojtak}, Rados{\l}aw and {Hansen}, Steen H. and {Hjorth}, Jens},
        title = "{Gravitational redshift of galaxies in clusters as predicted by general relativity}",
      journal = {\nat},
         year = 2011,
        month = sep,
       volume = {477},
       number = {7366},
        pages = {567-569},
          doi = {10.1038/nature10445},
archivePrefix = {arXiv},
       eprint = {1109.6571},
 primaryClass = {astro-ph.CO},
       adsurl = {https://ui.adsabs.harvard.edu/abs/2011Natur.477..567W}
}

@ARTICLE{1995A&A...301....6C,
       author = {{Cappi}, A.},
        title = "{Gravitational redshift in galaxy clusters.}",
      journal = {\aap},
         year = 1995,
        month = sep,
       volume = {301},
        pages = {6},
       adsurl = {https://ui.adsabs.harvard.edu/abs/1995A&A...301....6C}
}

@ARTICLE{2000PhLB..485..208D,
       author = {{Dvali}, G. and {Gabadadze}, G. and {Porrati}, M.},
        title = "{4D gravity on a brane in 5D Minkowski space}",
      journal = {Physics Letters B},
         year = 2000,
        month = jul,
       volume = {485},
       number = {1-3},
        pages = {208-214},
          doi = {10.1016/S0370-2693(00)00669-9},
archivePrefix = {arXiv},
       eprint = {hep-th/0005016},
 primaryClass = {hep-th},
       adsurl = {https://ui.adsabs.harvard.edu/abs/2000PhLB..485..208D}
}

@ARTICLE{2010RvMP...82..451S,
       author = {{Sotiriou}, Thomas P. and {Faraoni}, Valerio},
        title = "{f(R) theories of gravity}",
      journal = {Reviews of Modern Physics},
         year = 2010,
        month = jan,
       volume = {82},
       number = {1},
        pages = {451-497},
          doi = {10.1103/RevModPhys.82.451},
archivePrefix = {arXiv},
       eprint = {0805.1726},
 primaryClass = {gr-qc},
       adsurl = {https://ui.adsabs.harvard.edu/abs/2010RvMP...82..451S}
}

@ARTICLE{2007PhRvD..76f4004H,
       author = {{Hu}, Wayne and {Sawicki}, Ignacy},
        title = "{Models of f(R) cosmic acceleration that evade solar system tests}",
      journal = {\prd},
         year = 2007,
        month = sep,
       volume = {76},
       number = {6},
          eid = {064004},
        pages = {064004},
          doi = {10.1103/PhysRevD.76.064004},
archivePrefix = {arXiv},
       eprint = {0705.1158},
 primaryClass = {astro-ph},
       adsurl = {https://ui.adsabs.harvard.edu/abs/2007PhRvD..76f4004H}
}

@article{10.1098/rsta.2011.0291,
    author = {Skordis, Constantinos},
    title = {Modifications of gravity},
    journal = {Philosophical Transactions of the Royal Society A: Mathematical, Physical and Engineering Sciences},
    volume = {369},
    number = {1957},
    pages = {4962-4975},
    year = {2011},
    month = {12},
    issn = {1364-503X},
    doi = {10.1098/rsta.2011.0291},
    url = {https://doi.org/10.1098/rsta.2011.0291},
    eprint = {https://royalsocietypublishing.org/rsta/article-pdf/369/1957/4962/408209/rsta.2011.0291.pdf},
}

@ARTICLE{2010PhRvD..81j3002S,
       author = {{Schmidt}, Fabian},
        title = "{Dynamical masses in modified gravity}",
      journal = {\prd},
         year = 2010,
        month = may,
       volume = {81},
       number = {10},
          eid = {103002},
        pages = {103002},
          doi = {10.1103/PhysRevD.81.103002},
archivePrefix = {arXiv},
       eprint = {1003.0409},
 primaryClass = {astro-ph.CO},
       adsurl = {https://ui.adsabs.harvard.edu/abs/2010PhRvD..81j3002S}
}

@ARTICLE{2013MNRAS.434.3008C,
       author = {{Croft}, Rupert A.~C.},
        title = "{Gravitational redshifts from large-scale structure}",
      journal = {\mnras},
         year = 2013,
        month = oct,
       volume = {434},
       number = {4},
        pages = {3008-3017},
          doi = {10.1093/mnras/stt1223},
archivePrefix = {arXiv},
       eprint = {1304.4124},
 primaryClass = {astro-ph.CO},
       adsurl = {https://ui.adsabs.harvard.edu/abs/2013MNRAS.434.3008C}
}

@ARTICLE{2017arXiv170905756S,
       author = {{Sakuma}, Daiki and {Terukina}, Ayumu and {Yamamoto}, Kazuhiro and {Hikage}, Chiaki},
        title = "{Gravitational Redshifts in Clusters and Voids}",
      journal = {arXiv e-prints},
         year = 2017,
        month = sep,
          eid = {arXiv:1709.05756},
        pages = {arXiv:1709.05756},
          doi = {10.48550/arXiv.1709.05756},
archivePrefix = {arXiv},
       eprint = {1709.05756},
 primaryClass = {astro-ph.CO},
       adsurl = {https://ui.adsabs.harvard.edu/abs/2017arXiv170905756S}
}

@ARTICLE{2025JCAP...07..080D,
       author = {{Dio}, Enea Di and {Castello}, Sveva and {Bonvin}, Camille},
        title = "{Gravitational redshift from galaxy clusters {\textemdash} a relativistic approach}",
      journal = {\jcap},
         year = 2025,
        month = jul,
       volume = {2025},
       number = {7},
          eid = {080},
        pages = {080},
          doi = {10.1088/1475-7516/2025/07/080},
archivePrefix = {arXiv},
       eprint = {2503.11585},
 primaryClass = {astro-ph.CO},
       adsurl = {https://ui.adsabs.harvard.edu/abs/2025JCAP...07..080D}
}

@INCOLLECTION{1990LNP...360...29N,
       author = {{Nottale}, Laurent},
        title = "{Gravitational Redshifts and Lensing by Large Scale Structures}",
    booktitle = {Gravitational Lensing},
         year = 1990,
       editor = {{Mellier}, Yannick and {Fort}, Bernard and {Soucail}, Genevieve},
       volume = {360},
        pages = {29},
        publisher = {Gravitational Lensing},
          doi = {10.1007/BFb0009230},
       adsurl = {https://ui.adsabs.harvard.edu/abs/1990LNP...360...29N}
}

@ARTICLE{2004ApJ...607..164K,
       author = {{Kim}, Young-Rae and {Croft}, Rupert A.~C.},
        title = "{Gravitational Redshifts in Simulated Galaxy Clusters}",
      journal = {\apj},
         year = 2004,
        month = may,
       volume = {607},
       number = {1},
        pages = {164-174},
          doi = {10.1086/383218},
archivePrefix = {arXiv},
       eprint = {astro-ph/0402047},
 primaryClass = {astro-ph},
       adsurl = {https://ui.adsabs.harvard.edu/abs/2004ApJ...607..164K}
}

@ARTICLE{2024JCAP...05..003C,
       author = {{Castello}, Sveva and {Mancarella}, Michele and {Grimm}, Nastassia and {Sobral-Blanco}, Daniel and {Tutusaus}, Isaac and {Bonvin}, Camille},
        title = "{Gravitational redshift constraints on the effective theory of interacting dark energy}",
      journal = {\jcap},
         year = 2024,
        month = may,
       volume = {2024},
       number = {5},
          eid = {003},
        pages = {003},
          doi = {10.1088/1475-7516/2024/05/003},
archivePrefix = {arXiv},
       eprint = {2311.14425},
 primaryClass = {astro-ph.CO},
       adsurl = {https://ui.adsabs.harvard.edu/abs/2024JCAP...05..003C}
}

@ARTICLE{2024PhRvD.110j3523C,
       author = {{Castello}, Sveva and {Wang}, Zhuangfei and {Dam}, Lawrence and {Bonvin}, Camille and {Pogosian}, Levon},
        title = "{Disentangling modified gravity from a dark force with gravitational redshift}",
      journal = {\prd},
         year = 2024,
        month = nov,
       volume = {110},
       number = {10},
          eid = {103523},
        pages = {103523},
          doi = {10.1103/PhysRevD.110.103523},
archivePrefix = {arXiv},
       eprint = {2404.09379},
 primaryClass = {astro-ph.CO},
       adsurl = {https://ui.adsabs.harvard.edu/abs/2024PhRvD.110j3523C}
}

@ARTICLE{2000ApJ...533L..93B,
       author = {{Broadhurst}, Tom and {Scannapieco}, Evan},
        title = "{Detecting the Gravitational Redshift of Cluster Gas}",
      journal = {\apjl},
         year = 2000,
        month = apr,
       volume = {533},
       number = {2},
        pages = {L93-L97},
          doi = {10.1086/312630},
archivePrefix = {arXiv},
       eprint = {astro-ph/9912412},
 primaryClass = {astro-ph},
       adsurl = {https://ui.adsabs.harvard.edu/abs/2000ApJ...533L..93B}
}

@ARTICLE{2023A&A...669A..29R,
       author = {{Rosselli}, D. and {Marulli}, F. and {Veropalumbo}, A. and {Cimatti}, A. and {Moscardini}, L.},
        title = "{Testing general relativity: New measurements of gravitational redshift in galaxy clusters}",
      journal = {\aap},
         year = 2023,
        month = jan,
       volume = {669},
          eid = {A29},
        pages = {A29},
          doi = {10.1051/0004-6361/202244244},
archivePrefix = {arXiv},
       eprint = {2206.05313},
 primaryClass = {astro-ph.CO},
       adsurl = {https://ui.adsabs.harvard.edu/abs/2023A&A...669A..29R}
}

@ARTICLE{2015MNRAS.448.1999J,
       author = {{Jimeno}, Pablo and {Broadhurst}, Tom and {Coupon}, Jean and {Umetsu}, Keiichi and {Lazkoz}, Ruth},
        title = "{Comparing gravitational redshifts of SDSS galaxy clusters with the magnified redshift enhancement of background BOSS galaxies}",
      journal = {\mnras},
         year = 2015,
        month = apr,
       volume = {448},
       number = {3},
        pages = {1999-2012},
          doi = {10.1093/mnras/stv117},
archivePrefix = {arXiv},
       eprint = {1410.6050},
 primaryClass = {astro-ph.CO},
       adsurl = {https://ui.adsabs.harvard.edu/abs/2015MNRAS.448.1999J}
}

@ARTICLE{2015PhRvL.114g1103S,
       author = {{Sadeh}, Iftach and {Feng}, Low Lerh and {Lahav}, Ofer},
        title = "{Gravitational Redshift of Galaxies in Clusters from the Sloan Digital Sky Survey and the Baryon Oscillation Spectroscopic Survey}",
      journal = {\prl},
         year = 2015,
        month = feb,
       volume = {114},
       number = {7},
          eid = {071103},
        pages = {071103},
          doi = {10.1103/PhysRevLett.114.071103},
archivePrefix = {arXiv},
       eprint = {1410.5262},
 primaryClass = {astro-ph.CO},
       adsurl = {https://ui.adsabs.harvard.edu/abs/2015PhRvL.114g1103S}
}

@ARTICLE{2025arXiv251213221T,
       author = {{Tsaprazi}, Eleni and {Lesci}, Giorgio F. and {Marulli}, Federico and {Heavens}, Alan F. and {Rosati}, Piero and {Contarini}, Sofia and {Maraboli}, Enrico A. and {Dayal}, Pratika and {Lahav}, Ofer and {Moscardini}, Lauro},
        title = "{Beyond general relativity: probing gravity with gravitational redshifts}",
      journal = {arXiv e-prints},
         year = 2025,
        month = dec,
          eid = {arXiv:2512.13221},
        pages = {arXiv:2512.13221},
          doi = {10.48550/arXiv.2512.13221},
archivePrefix = {arXiv},
       eprint = {2512.13221},
 primaryClass = {astro-ph.CO},
       adsurl = {https://ui.adsabs.harvard.edu/abs/2025arXiv251213221T}
}

@ARTICLE{2019IJMPD..2850150X,
       author = {{Xu}, Haoting and {Huang}, Zhiqi and {Zhang}, Na and {Jiang}, Yundong},
        title = "{Forecasting cosmological bias due to local gravitational redshift}",
      journal = {International Journal of Modern Physics D},
         year = 2019,
        month = jan,
       volume = {28},
       number = {12},
          eid = {1950150},
        pages = {1950150},
          doi = {10.1142/S0218271819501505},
archivePrefix = {arXiv},
       eprint = {1904.01753},
 primaryClass = {astro-ph.CO},
       adsurl = {https://ui.adsabs.harvard.edu/abs/2019IJMPD..2850150X}
}

@ARTICLE{2022MNRAS.511.2732S,
       author = {{Saga}, Shohei and {Taruya}, Atsushi and {Rasera}, Yann and {Breton}, Michel-Andr{\`e}s},
        title = "{Detectability of the gravitational redshift effect from the asymmetric galaxy clustering}",
      journal = {\mnras},
         year = 2022,
        month = apr,
       volume = {511},
       number = {2},
        pages = {2732-2754},
          doi = {10.1093/mnras/stac186},
archivePrefix = {arXiv},
       eprint = {2109.06012},
 primaryClass = {astro-ph.CO},
       adsurl = {https://ui.adsabs.harvard.edu/abs/2022MNRAS.511.2732S}
}

@ARTICLE{2025A&A...697A...1E,
       author = {{Euclid Collaboration} and {Mellier}, Y. and {Abdurro'uf} and {Acevedo Barroso}, J.~A. and {Ach{\'u}carro}, A. and {Adamek}, J. and {Adam}, R. and {Addison}, G.~E. and {Aghanim}, N. and {Aguena}, M. and {Ajani}, V. and {Akrami}, Y. and {Al-Bahlawan}, A. and {Alavi}, A. and {Albuquerque}, I.~S. and {Alestas}, G. and {Alguero}, G. and {Allaoui}, A. and {Allen}, S.~W. and {Allevato}, V. and {Alonso-Tetilla}, A.~V. and {Altieri}, B. and {Alvarez-Candal}, A. and {Alvi}, S. and {Amara}, A. and {Amendola}, L. and {Amiaux}, J. and {Andika}, I.~T. and {Andreon}, S. and {Andrews}, A. and {Angora}, G. and {Angulo}, R.~E. and {Annibali}, F. and {Anselmi}, A. and {Anselmi}, S. and {Arcari}, S. and {Archidiacono}, M. and {Aric{\`o}}, G. and {Arnaud}, M. and {Arnouts}, S. and {Asgari}, M. and {Asorey}, J. and {Atayde}, L. and {Atek}, H. and {Atrio-Barandela}, F. and {Aubert}, M. and {Aubourg}, E. and {Auphan}, T. and {Auricchio}, N. and {Aussel}, B. and {Aussel}, H. and {Avelino}, P.~P. and {Avgoustidis}, A. and {Avila}, S. and {Awan}, S. and {Azzollini}, R. and {Baccigalupi}, C. and {Bachelet}, E. and {Bacon}, D. and {Baes}, M. and {Bagley}, M.~B. and {Bahr-Kalus}, B. and {Balaguera-Antolinez}, A. and {Balbinot}, E. and {Balcells}, M. and {Baldi}, M. and {Baldry}, I. and {Balestra}, A. and {Ballardini}, M. and {Ballester}, O. and {Balogh}, M. and {Ba{\~n}ados}, E. and {Barbier}, R. and {Bardelli}, S. and {Baron}, M. and {Barreiro}, T. and {Barrena}, R. and {Barriere}, J.-C. and {Barros}, B.~J. and {Barthelemy}, A. and {Bartolo}, N. and {Basset}, A. and {Battaglia}, P. and {Battisti}, A.~J. and {Baugh}, C.~M. and {Baumont}, L. and {Bazzanini}, L. and {Beaulieu}, J.-P. and {Beckmann}, V. and {Belikov}, A.~N. and {Bel}, J. and {Bellagamba}, F. and {Bella}, M. and {Bellini}, E. and {Benabed}, K. and {Bender}, R. and {Benevento}, G. and {Bennett}, C.~L. and {Benson}, K. and {Bergamini}, P. and {Bermejo-Climent}, J.~R. and {Bernardeau}, F. and {Bertacca}, D. and {Berthe}, M. and {Berthier}, J. and {Bethermin}, M. and {Beutler}, F. and {Bevillon}, C. and {Bhargava}, S. and {Bhatawdekar}, R. and {Bianchi}, D. and {Bisigello}, L. and {Biviano}, A. and {Blake}, R.~P. and {Blanchard}, A. and {Blazek}, J. and {Blot}, L. and {Bosco}, A. and {Bodendorf}, C. and {Boenke}, T. and {B{\"o}hringer}, H. and {Boldrini}, P. and {Bolzonella}, M. and {Bonchi}, A. and {Bonici}, M. and {Bonino}, D. and {Bonino}, L. and {Bonvin}, C. and {Bon}, W. and {Booth}, J.~T. and {Borgani}, S. and {Borlaff}, A.~S. and {Borsato}, E. and {Bose}, B. and {Botticella}, M.~T. and {Boucaud}, A. and {Bouche}, F. and {Boucher}, J.~S. and {Boutigny}, D. and {Bouvard}, T. and {Bouwens}, R. and {Bouy}, H. and {Bowler}, R.~A.~A. and {Bozza}, V. and {Bozzo}, E. and {Branchini}, E. and {Brando}, G. and {Brau-Nogue}, S. and {Brekke}, P. and {Bremer}, M.~N. and {Brescia}, M. and {Breton}, M.-A. and {Brinchmann}, J. and {Brinckmann}, T. and {Brockley-Blatt}, C. and {Brodwin}, M. and {Brouard}, L. and {Brown}, M.~L. and {Bruton}, S. and {Bucko}, J. and {Buddelmeijer}, H. and {Buenadicha}, G. and {Buitrago}, F. and {Burger}, P. and {Burigana}, C. and {Busillo}, V. and {Busonero}, D. and {Cabanac}, R. and {Cabayol-Garcia}, L. and {Cagliari}, M.~S. and {Caillat}, A. and {Caillat}, L. and {Calabrese}, M. and {Calabro}, A. and {Calderone}, G. and {Calura}, F. and {Camacho Quevedo}, B. and {Camera}, S. and {Campos}, L. and {Ca{\~n}as-Herrera}, G. and {Candini}, G.~P. and {Cantiello}, M. and {Capobianco}, V. and {Cappellaro}, E. and {Cappelluti}, N. and {Cappi}, A. and {Caputi}, K.~I. and {Cara}, C. and {Carbone}, C. and {Cardone}, V.~F. and {Carella}, E. and {Carlberg}, R.~G. and {Carle}, M. and {Carminati}, L. and {Caro}, F. and {Carrasco}, J.~M. and {Carretero}, J. and {Carrilho}, P. and {Carron Duque}, J. and {Carry}, B.},
        title = "{Euclid: I. Overview of the Euclid mission}",
      journal = {\aap},
         year = 2025,
        month = may,
       volume = {697},
          eid = {A1},
        pages = {A1},
          doi = {10.1051/0004-6361/202450810},
archivePrefix = {arXiv},
       eprint = {2405.13491},
 primaryClass = {astro-ph.CO},
       adsurl = {https://ui.adsabs.harvard.edu/abs/2025A&A...697A...1E}
}

@ARTICLE{2024ApJS..272...39W,
       author = {{Wen}, Z.~L. and {Han}, J.~L.},
        title = "{A Catalog of 1.58 Million Clusters of Galaxies Identified from the DESI Legacy Imaging Surveys}",
      journal = {\apjs},
         year = 2024,
        month = jun,
       volume = {272},
       number = {2},
          eid = {39},
        pages = {39},
          doi = {10.3847/1538-4365/ad409d},
archivePrefix = {arXiv},
       eprint = {2404.02002},
 primaryClass = {astro-ph.CO},
       adsurl = {https://ui.adsabs.harvard.edu/abs/2024ApJS..272...39W}
}

@ARTICLE{2009arXiv09120201L,
       author = {{LSST Science Collaboration} and {Abell}, Paul A. and {Allison}, Julius and {Anderson}, Scott F. and {Andrew}, John R. and {Angel}, J. Roger P. and {Armus}, Lee and {Arnett}, David and {Asztalos}, S.~J. and {Axelrod}, Tim S. and {Bailey}, Stephen and {Ballantyne}, D.~R. and {Bankert}, Justin R. and {Barkhouse}, Wayne A. and {Barr}, Jeffrey D. and {Barrientos}, L. Felipe and {Barth}, Aaron J. and {Bartlett}, James G. and {Becker}, Andrew C. and {Becla}, Jacek and {Beers}, Timothy C. and {Bernstein}, Joseph P. and {Biswas}, Rahul and {Blanton}, Michael R. and {Bloom}, Joshua S. and {Bochanski}, John J. and {Boeshaar}, Pat and {Borne}, Kirk D. and {Bradac}, Marusa and {Brandt}, W.~N. and {Bridge}, Carrie R. and {Brown}, Michael E. and {Brunner}, Robert J. and {Bullock}, James S. and {Burgasser}, Adam J. and {Burge}, James H. and {Burke}, David L. and {Cargile}, Phillip A. and {Chandrasekharan}, Srinivasan and {Chartas}, George and {Chesley}, Steven R. and {Chu}, You-Hua and {Cinabro}, David and {Claire}, Mark W. and {Claver}, Charles F. and {Clowe}, Douglas and {Connolly}, A.~J. and {Cook}, Kem H. and {Cooke}, Jeff and {Cooray}, Asantha and {Covey}, Kevin R. and {Culliton}, Christopher S. and {de Jong}, Roelof and {de Vries}, Willem H. and {Debattista}, Victor P. and {Delgado}, Francisco and {Dell'Antonio}, Ian P. and {Dhital}, Saurav and {Di Stefano}, Rosanne and {Dickinson}, Mark and {Dilday}, Benjamin and {Djorgovski}, S.~G. and {Dobler}, Gregory and {Donalek}, Ciro and {Dubois-Felsmann}, Gregory and {Durech}, Josef and {Eliasdottir}, Ardis and {Eracleous}, Michael and {Eyer}, Laurent and {Falco}, Emilio E. and {Fan}, Xiaohui and {Fassnacht}, Christopher D. and {Ferguson}, Harry C. and {Fernandez}, Yanga R. and {Fields}, Brian D. and {Finkbeiner}, Douglas and {Figueroa}, Eduardo E. and {Fox}, Derek B. and {Francke}, Harold and {Frank}, James S. and {Frieman}, Josh and {Fromenteau}, Sebastien and {Furqan}, Muhammad and {Galaz}, Gaspar and {Gal-Yam}, A. and {Garnavich}, Peter and {Gawiser}, Eric and {Geary}, John and {Gee}, Perry and {Gibson}, Robert R. and {Gilmore}, Kirk and {Grace}, Emily A. and {Green}, Richard F. and {Gressler}, William J. and {Grillmair}, Carl J. and {Habib}, Salman and {Haggerty}, J.~S. and {Hamuy}, Mario and {Harris}, Alan W. and {Hawley}, Suzanne L. and {Heavens}, Alan F. and {Hebb}, Leslie and {Henry}, Todd J. and {Hileman}, Edward and {Hilton}, Eric J. and {Hoadley}, Keri and {Holberg}, J.~B. and {Holman}, Matt J. and {Howell}, Steve B. and {Infante}, Leopoldo and {Ivezic}, Zeljko and {Jacoby}, Suzanne H. and {Jain}, Bhuvnesh and {R} and {Jedicke} and {Jee}, M. James and {Garrett Jernigan}, J. and {Jha}, Saurabh W. and {Johnston}, Kathryn V. and {Jones}, R. Lynne and {Juric}, Mario and {Kaasalainen}, Mikko and {Styliani} and {Kafka} and {Kahn}, Steven M. and {Kaib}, Nathan A. and {Kalirai}, Jason and {Kantor}, Jeff and {Kasliwal}, Mansi M. and {Keeton}, Charles R. and {Kessler}, Richard and {Knezevic}, Zoran and {Kowalski}, Adam and {Krabbendam}, Victor L. and {Krughoff}, K. Simon and {Kulkarni}, Shrinivas and {Kuhlman}, Stephen and {Lacy}, Mark and {Lepine}, Sebastien and {Liang}, Ming and {Lien}, Amy and {Lira}, Paulina and {Long}, Knox S. and {Lorenz}, Suzanne and {Lotz}, Jennifer M. and {Lupton}, R.~H. and {Lutz}, Julie and {Macri}, Lucas M. and {Mahabal}, Ashish A. and {Mandelbaum}, Rachel and {Marshall}, Phil and {May}, Morgan and {McGehee}, Peregrine M. and {Meadows}, Brian T. and {Meert}, Alan and {Milani}, Andrea and {Miller}, Christopher J. and {Miller}, Michelle and {Mills}, David and {Minniti}, Dante and {Monet}, David and {Mukadam}, Anjum S. and {Nakar}, Ehud and {Neill}, Douglas R. and {Newman}, Jeffrey A. and {Nikolaev}, Sergei and {Nordby}, Martin and {O'Connor}, Paul and {Oguri}, Masamune and {Oliver}, John and {Olivier}, Scot S. and {Olsen}, Julia K. and {Olsen}, Knut and {Olszewski}, Edward W. and {Oluseyi}, Hakeem and {Padilla}, Nelson D. and {Parker}, Alex and {Pepper}, Joshua and {Peterson}, John R. and {Petry}, Catherine and {Pinto}, Philip A. and {Pizagno}, James L. and {Popescu}, Bogdan and {Prsa}, Andrej and {Radcka}, Veljko and {Raddick}, M. Jordan and {Rasmussen}, Andrew and {Rau}, Arne and {Rho}, Jeonghee and {Rhoads}, James E. and {Richards}, Gordon T. and {Ridgway}, Stephen T. and {Robertson}, Brant E. and {Roskar}, Rok and {Saha}, Abhijit and {Sarajedini}, Ata and {Scannapieco}, Evan and {Schalk}, Terry and {Schindler}, Rafe and {Schmidt}, Samuel},
        title = "{LSST Science Book, Version 2.0}",
      journal = {arXiv e-prints},
         year = 2009,
        month = dec,
          eid = {arXiv:0912.0201},
        pages = {arXiv:0912.0201},
          doi = {10.48550/arXiv.0912.0201},
archivePrefix = {arXiv},
       eprint = {0912.0201},
 primaryClass = {astro-ph.IM},
       adsurl = {https://ui.adsabs.harvard.edu/abs/2009arXiv0912.0201L}
}

@ARTICLE{2005ApJ...633..791Z,
       author = {{Zheng}, Zheng and {Berlind}, Andreas A. and {Weinberg}, David H. and {Benson}, Andrew J. and {Baugh}, Carlton M. and {Cole}, Shaun and {Dav{\'e}}, Romeel and {Frenk}, Carlos S. and {Katz}, Neal and {Lacey}, Cedric G.},
        title = "{Theoretical Models of the Halo Occupation Distribution: Separating Central and Satellite Galaxies}",
      journal = {\apj},
         year = 2005,
        month = nov,
       volume = {633},
       number = {2},
        pages = {791-809},
          doi = {10.1086/466510},
archivePrefix = {arXiv},
       eprint = {astro-ph/0408564},
 primaryClass = {astro-ph},
       adsurl = {https://ui.adsabs.harvard.edu/abs/2005ApJ...633..791Z}
}

@ARTICLE{1997ApJ...490..493N,
       author = {{Navarro}, Julio F. and {Frenk}, Carlos S. and {White}, Simon D.~M.},
        title = "{A Universal Density Profile from Hierarchical Clustering}",
      journal = {\apj},
         year = 1997,
        month = dec,
       volume = {490},
       number = {2},
        pages = {493-508},
          doi = {10.1086/304888},
archivePrefix = {arXiv},
       eprint = {astro-ph/9611107},
 primaryClass = {astro-ph},
       adsurl = {https://ui.adsabs.harvard.edu/abs/1997ApJ...490..493N}
}

@ARTICLE{2001MNRAS.322..901S,
       author = {{Sheth}, Ravi K. and {Diaferio}, Antonaldo},
        title = "{Peculiar velocities of galaxies and clusters}",
      journal = {\mnras},
         year = 2001,
        month = apr,
       volume = {322},
       number = {4},
        pages = {901-917},
          doi = {10.1046/j.1365-8711.2001.04202.x},
archivePrefix = {arXiv},
       eprint = {astro-ph/0009166},
 primaryClass = {astro-ph},
       adsurl = {https://ui.adsabs.harvard.edu/abs/2001MNRAS.322..901S}
}

@ARTICLE{2009ApJ...696.2115I,
       author = {{Ishiyama}, Tomoaki and {Fukushige}, Toshiyuki and {Makino}, Junichiro},
        title = "{Variation of the Subhalo Abundance in Dark Matter Halos}",
      journal = {\apj},
         year = 2009,
        month = may,
       volume = {696},
       number = {2},
        pages = {2115-2125},
          doi = {10.1088/0004-637X/696/2/2115},
archivePrefix = {arXiv},
       eprint = {0812.0683},
 primaryClass = {astro-ph},
       adsurl = {https://ui.adsabs.harvard.edu/abs/2009ApJ...696.2115I}
}

@ARTICLE{2020A&A...641A...6P,
       author = {{Planck Collaboration} and {Aghanim}, N. and {Akrami}, Y. and {Ashdown}, M. and {Aumont}, J. and {Baccigalupi}, C. and {Ballardini}, M. and {Banday}, A.~J. and {Barreiro}, R.~B. and {Bartolo}, N. and {Basak}, S. and {Battye}, R. and {Benabed}, K. and {Bernard}, J.-P. and {Bersanelli}, M. and {Bielewicz}, P. and {Bock}, J.~J. and {Bond}, J.~R. and {Borrill}, J. and {Bouchet}, F.~R. and {Boulanger}, F. and {Bucher}, M. and {Burigana}, C. and {Butler}, R.~C. and {Calabrese}, E. and {Cardoso}, J.-F. and {Carron}, J. and {Challinor}, A. and {Chiang}, H.~C. and {Chluba}, J. and {Colombo}, L.~P.~L. and {Combet}, C. and {Contreras}, D. and {Crill}, B.~P. and {Cuttaia}, F. and {de Bernardis}, P. and {de Zotti}, G. and {Delabrouille}, J. and {Delouis}, J.-M. and {Di Valentino}, E. and {Diego}, J.~M. and {Dor{\'e}}, O. and {Douspis}, M. and {Ducout}, A. and {Dupac}, X. and {Dusini}, S. and {Efstathiou}, G. and {Elsner}, F. and {En{\ss}lin}, T.~A. and {Eriksen}, H.~K. and {Fantaye}, Y. and {Farhang}, M. and {Fergusson}, J. and {Fernandez-Cobos}, R. and {Finelli}, F. and {Forastieri}, F. and {Frailis}, M. and {Fraisse}, A.~A. and {Franceschi}, E. and {Frolov}, A. and {Galeotta}, S. and {Galli}, S. and {Ganga}, K. and {G{\'e}nova-Santos}, R.~T. and {Gerbino}, M. and {Ghosh}, T. and {Gonz{\'a}lez-Nuevo}, J. and {G{\'o}rski}, K.~M. and {Gratton}, S. and {Gruppuso}, A. and {Gudmundsson}, J.~E. and {Hamann}, J. and {Handley}, W. and {Hansen}, F.~K. and {Herranz}, D. and {Hildebrandt}, S.~R. and {Hivon}, E. and {Huang}, Z. and {Jaffe}, A.~H. and {Jones}, W.~C. and {Karakci}, A. and {Keih{\"a}nen}, E. and {Keskitalo}, R. and {Kiiveri}, K. and {Kim}, J. and {Kisner}, T.~S. and {Knox}, L. and {Krachmalnicoff}, N. and {Kunz}, M. and {Kurki-Suonio}, H. and {Lagache}, G. and {Lamarre}, J.-M. and {Lasenby}, A. and {Lattanzi}, M. and {Lawrence}, C.~R. and {Le Jeune}, M. and {Lemos}, P. and {Lesgourgues}, J. and {Levrier}, F. and {Lewis}, A. and {Liguori}, M. and {Lilje}, P.~B. and {Lilley}, M. and {Lindholm}, V. and {L{\'o}pez-Caniego}, M. and {Lubin}, P.~M. and {Ma}, Y.-Z. and {Mac{\'\i}as-P{\'e}rez}, J.~F. and {Maggio}, G. and {Maino}, D. and {Mandolesi}, N. and {Mangilli}, A. and {Marcos-Caballero}, A. and {Maris}, M. and {Martin}, P.~G. and {Martinelli}, M. and {Mart{\'\i}nez-Gonz{\'a}lez}, E. and {Matarrese}, S. and {Mauri}, N. and {McEwen}, J.~D. and {Meinhold}, P.~R. and {Melchiorri}, A. and {Mennella}, A. and {Migliaccio}, M. and {Millea}, M. and {Mitra}, S. and {Miville-Desch{\^e}nes}, M.-A. and {Molinari}, D. and {Montier}, L. and {Morgante}, G. and {Moss}, A. and {Natoli}, P. and {N{\o}rgaard-Nielsen}, H.~U. and {Pagano}, L. and {Paoletti}, D. and {Partridge}, B. and {Patanchon}, G. and {Peiris}, H.~V. and {Perrotta}, F. and {Pettorino}, V. and {Piacentini}, F. and {Polastri}, L. and {Polenta}, G. and {Puget}, J.-L. and {Rachen}, J.~P. and {Reinecke}, M. and {Remazeilles}, M. and {Renzi}, A. and {Rocha}, G. and {Rosset}, C. and {Roudier}, G. and {Rubi{\~n}o-Mart{\'\i}n}, J.~A. and {Ruiz-Granados}, B. and {Salvati}, L. and {Sandri}, M. and {Savelainen}, M. and {Scott}, D. and {Shellard}, E.~P.~S. and {Sirignano}, C. and {Sirri}, G. and {Spencer}, L.~D. and {Sunyaev}, R. and {Suur-Uski}, A.-S. and {Tauber}, J.~A. and {Tavagnacco}, D. and {Tenti}, M. and {Toffolatti}, L. and {Tomasi}, M. and {Trombetti}, T. and {Valenziano}, L. and {Valiviita}, J. and {Van Tent}, B. and {Vibert}, L. and {Vielva}, P. and {Villa}, F. and {Vittorio}, N. and {Wandelt}, B.~D. and {Wehus}, I.~K. and {White}, M. and {White}, S.~D.~M. and {Zacchei}, A. and {Zonca}, A.},
        title = "{Planck 2018 results. VI. Cosmological parameters}",
      journal = {\aap},
         year = 2020,
        month = sep,
       volume = {641},
          eid = {A6},
        pages = {A6},
          doi = {10.1051/0004-6361/201833910},
archivePrefix = {arXiv},
       eprint = {1807.06209},
 primaryClass = {astro-ph.CO},
       adsurl = {https://ui.adsabs.harvard.edu/abs/2020A&A...641A...6P}
}

@ARTICLE{2024arXiv240305398M,
       author = {{Mainieri}, Vincenzo and {Anderson}, Richard I. and {Brinchmann}, Jarle and {Cimatti}, Andrea and {Ellis}, Richard S. and {Hill}, Vanessa and {Kneib}, Jean-Paul and {McLeod}, Anna F. and {Opitom}, Cyrielle and {Roth}, Martin M. and {Sanchez-Saez}, Paula and {Smiljanic}, Rodolfo and {Tolstoy}, Eline and {Bacon}, Roland and {Randich}, Sofia and {Adamo}, Angela and {Annibali}, Francesca and {Arevalo}, Patricia and {Audard}, Marc and {Barsanti}, Stefania and {Battaglia}, Giuseppina and {Bayo Aran}, Amelia M. and {Belfiore}, Francesco and {Bellazzini}, Michele and {Bellini}, Emilio and {Beltran}, Maria Teresa and {Berni}, Leda and {Bianchi}, Simone and {Biazzo}, Katia and {Bisero}, Sofia and {Bisogni}, Susanna and {Bland-Hawthorn}, Joss and {Blondin}, Stephane and {Bodensteiner}, Julia and {Boffin}, Henri M.~J. and {Bonito}, Rosaria and {Bono}, Giuseppe and {Bouche}, Nicolas F. and {Bowman}, Dominic and {Braga}, Vittorio F. and {Bragaglia}, Angela and {Branchesi}, Marica and {Brucalassi}, Anna and {Bryant}, Julia J. and {Bryson}, Ian and {Busa}, Innocenza and {Camera}, Stefano and {Carbone}, Carmelita and {Casali}, Giada and {Casali}, Mark and {Casasola}, Viviana and {Castro}, Norberto and {Catelan}, Marcio and {Cavallo}, Lorenzo and {Chiappini}, Cristina and {Cioni}, Maria-Rosa and {Colless}, Matthew and {Colzi}, Laura and {Contarini}, Sofia and {Couch}, Warrick and {D'Ammando}, Filippo and {d'Assignies D.}, William and {D'Orazi}, Valentina and {da Silva}, Ronaldo and {Dainotti}, Maria Giovanna and {Damiani}, Francesco and {Danielski}, Camilla and {De Cia}, Annalisa and {de Jong}, Roelof S. and {Dhawan}, Suhail and {Dierickx}, Philippe and {Driver}, Simon P. and {Dupletsa}, Ulyana and {Escoffier}, Stephanie and {Escorza}, Ana and {Fabrizio}, Michele and {Fiorentino}, Giuliana and {Fontana}, Adriano and {Fontani}, Francesco and {Forero Sanchez}, Daniel and {Franois}, Patrick and {Galindo-Guil}, Francisco Jose and {Gallazzi}, Anna Rita and {Galli}, Daniele and {Garcia}, Miriam and {Garcia-Rojas}, Jorge and {Garilli}, Bianca and {Grand}, Robert and {Guarcello}, Mario Giuseppe and {Hazra}, Nandini and {Helmi}, Amina and {Herrero}, Artemio and {Iglesias}, Daniela and {Ilic}, Dragana and {Irsic}, Vid and {Ivanov}, Valentin D. and {Izzo}, Luca and {Jablonka}, Pascale and {Joachimi}, Benjamin and {Kakkad}, Darshan and {Kamann}, Sebastian and {Koposov}, Sergey and {Kordopatis}, Georges and {Kovacevic}, Andjelka B. and {Kraljic}, Katarina and {Kuncarayakti}, Hanindyo and {Kwon}, Yuna and {La Forgia}, Fiorangela and {Lahav}, Ofer and {Laigle}, Clotilde and {Lazzarin}, Monica and {Leaman}, Ryan and {Leclercq}, Floriane and {Lee}, Khee-Gan and {Lee}, David and {Lehnert}, Matt D. and {Lira}, Paulina and {Loffredo}, Eleonora and {Lucatello}, Sara and {Magrini}, Laura and {Maguire}, Kate and {Mahler}, Guillaume and {Zahra Majidi}, Fatemeh and {Malavasi}, Nicola and {Mannucci}, Filippo and {Marconi}, Marcella and {Martin}, Nicolas and {Marulli}, Federico and {Massari}, Davide and {Matsuno}, Tadafumi and {Mattheee}, Jorryt and {McGee}, Sean and {Merc}, Jaroslav and {Merle}, Thibault and {Miglio}, Andrea and {Migliorini}, Alessandra and {Minchev}, Ivan and {Minniti}, Dante and {Miret-Roig}, Nuria and {Monreal Ibero}, Ana and {Montano}, Federico and {Montet}, Ben T. and {Moresco}, Michele and {Moretti}, Chiara and {Moscardini}, Lauro and {Moya}, Andres and {Mueller}, Oliver and {Nanayakkara}, Themiya and {Nicholl}, Matt and {Nordlander}, Thomas and {Onori}, Francesca and {Padovani}, Marco and {Pala}, Anna Francesca and {Panda}, Swayamtrupta and {Pandey-Pommier}, Mamta and {Pasquini}, Luca and {Pawlak}, Michal and {Pessi}, Priscila J. and {Pisani}, Alice and {Popovic}, Lukav C. and {Prisinzano}, Loredana and {Raddi}, Roberto and {Rainer}, Monica and {Rebassa-Mansergas}, Alberto and {Richard}, Johan and {Rigault}, Mickael and {Rocher}, Antoine and {Romano}, Donatella and {Rosati}, Piero and {Sacco}, Germano and {Sanchez-Janssen}, Ruben and {Sander}, Andreas A.~C. and {Sanders}, Jason L. and {Sargent}, Mark and {Sarpa}, Elena and {Schimd}, Carlo and {Schipani}, Pietro and {Sefusatti}, Emiliano and {Smith}, Graham P. and {Spina}, Lorenzo and {Steinmetz}, Matthias and {Tacchella}, Sandro and {Tautvaisiene}, Grazina and {Theissen}, Christopher and {Thomas}, Guillaume and {Ting}, Yuan-Sen and {Travouillon}, Tony and {Tresse}, Laurence and {Trivedi}, Oem and {Tsantaki}, Maria and {Tsedrik}, Maria and {Urrutia}, Tanya and {Valenti}, Elena and {Van der Swaelmen}, Mathieu and {Van Eck}, Sophie and {Verdiani}, Francesco and {Verdier}, Aurelien and {Vergani}, Susanna Diana and {Verhamme}, Anne and {Vernet}, Joel},
        title = "{The Wide-field Spectroscopic Telescope (WST) Science White Paper}",
      journal = {arXiv e-prints},
         year = 2024,
        month = mar,
          eid = {arXiv:2403.05398},
        pages = {arXiv:2403.05398},
          doi = {10.48550/arXiv.2403.05398},
archivePrefix = {arXiv},
       eprint = {2403.05398},
 primaryClass = {astro-ph.IM},
       adsurl = {https://ui.adsabs.harvard.edu/abs/2024arXiv240305398M}
}

@ARTICLE{2022JCAP...05..018R,
       author = {{Ruan}, Cheng-Zong and {Hern{\'a}ndez-Aguayo}, C{\'e}sar and {Li}, Baojiu and {Arnold}, Christian and {Baugh}, Carlton M. and {Klypin}, Anatoly and {Prada}, Francisco},
        title = "{Fast full N-body simulations of generic modified gravity: conformal coupling models}",
      journal = {\jcap},
         year = 2022,
        month = may,
       volume = {2022},
       number = {5},
          eid = {018},
        pages = {018},
          doi = {10.1088/1475-7516/2022/05/018},
archivePrefix = {arXiv},
       eprint = {2110.00328},
 primaryClass = {astro-ph.CO},
       adsurl = {https://ui.adsabs.harvard.edu/abs/2022JCAP...05..018R}
}

@ARTICLE{2018MNRAS.481.2813G,
       author = {{Giocoli}, Carlo and {Baldi}, Marco and {Moscardini}, Lauro},
        title = "{Weak lensing light-cones in modified gravity simulations with and without massive neutrinos}",
      journal = {\mnras},
         year = 2018,
        month = dec,
       volume = {481},
       number = {2},
        pages = {2813-2828},
          doi = {10.1093/mnras/sty2465},
archivePrefix = {arXiv},
       eprint = {1806.04681},
 primaryClass = {astro-ph.CO},
       adsurl = {https://ui.adsabs.harvard.edu/abs/2018MNRAS.481.2813G}
}

@ARTICLE{2021JCAP...09..024B,
       author = {{Brando}, Guilherme and {Koyama}, Kazuya and {Wands}, David and {Zumalac{\'a}rregui}, Miguel and {Sawicki}, Ignacy and {Bellini}, Emilio},
        title = "{Fully relativistic predictions in Horndeski gravity from standard Newtonian N-body simulations}",
      journal = {\jcap},
         year = 2021,
        month = sep,
       volume = {2021},
       number = {9},
          eid = {024},
        pages = {024},
          doi = {10.1088/1475-7516/2021/09/024},
archivePrefix = {arXiv},
       eprint = {2105.04491},
 primaryClass = {astro-ph.CO},
       adsurl = {https://ui.adsabs.harvard.edu/abs/2021JCAP...09..024B}
}

@ARTICLE{18,
       author = {{Christodoulou} and {Kazanas}, Demosthenes},
        title = "{Gravitational potential and non-relativistic Lagrangian in modified gravity with varying G}",
      journal = {MNRAS},
         year = 2019,
        month = feb,
       volume = {483},
       number = {1},
        eid = {L85-L87},
          doi = {10.1093/mnrasl/sly222},
       adsurl = {https://ui.adsabs.harvard.edu/abs/2019MNRAS.483L..85C}
}

@ARTICLE{2022PDU....3701069J,
       author = {{Jaber}, Mariana and {Arciniega}, Gustavo and {Jaime}, Luisa G. and {Rodr{\'\i}guez-L{\'o}pez}, Omar Abel},
        title = "{A single parameterization for dark energy and modified gravity models}",
      journal = {Physics of the Dark Universe},
         year = 2022,
        month = sep,
       volume = {37},
          eid = {101069},
        pages = {101069},
          doi = {10.1016/j.dark.2022.101069},
archivePrefix = {arXiv},
       eprint = {2102.08561},
 primaryClass = {astro-ph.CO},
       adsurl = {https://ui.adsabs.harvard.edu/abs/2022PDU....3701069J}
}

@ARTICLE{2013PhRvD..88h3513N,
       author = {{Narimani}, Ali and {Scott}, Douglas},
        title = "{Minimal parameterizations for modified gravity}",
      journal = {\prd},
         year = 2013,
        month = oct,
       volume = {88},
       number = {8},
          eid = {083513},
        pages = {083513},
          doi = {10.1103/PhysRevD.88.083513},
archivePrefix = {arXiv},
       eprint = {1303.3197},
 primaryClass = {astro-ph.CO},
       adsurl = {https://ui.adsabs.harvard.edu/abs/2013PhRvD..88h3513N}
}

@ARTICLE{Tinker08,
       author = {{Tinker}, Jeremy and {Kravtsov}, Andrey V. and {Klypin}, Anatoly and {Abazajian}, Kevork and {Warren}, Michael and {Yepes}, Gustavo and {Gottl{\"o}ber}, Stefan and {Holz}, Daniel E.},
        title = "{Toward a Halo Mass Function for Precision Cosmology: The Limits of Universality}",
      journal = {\apj},
         year = 2008,
        month = dec,
       volume = {688},
       number = {2},
        pages = {709-728},
          doi = {10.1086/591439},
archivePrefix = {arXiv},
       eprint = {0803.2706},
 primaryClass = {astro-ph},
       adsurl = {https://ui.adsabs.harvard.edu/abs/2008ApJ...688..709T}
}

@ARTICLE{Mamon_2013_MAMPOSSt,
       author = {{Mamon}, Gary A. and {Biviano}, Andrea and {Bou{\'e}}, Gwena{\"e}l},
        title = "{MAMPOSSt: Modelling Anisotropy and Mass Profiles of Observed Spherical Systems - I. Gaussian 3D velocities}",
      journal = {MNRAS},
         year = 2013,
        month = mar,
       volume = {429},
       number = {4},
        pages = {3079-3098},
          doi = {10.1093/mnras/sts565},
archivePrefix = {arXiv},
       eprint = {1212.1455},
 primaryClass = {astro-ph.CO},
       adsurl = {https://ui.adsabs.harvard.edu/abs/2013MNRAS.429.3079M}
}

@ARTICLE{AguirreTagliaferro_2021_MAMPOSStvalidationNbodysim,
       author = {{Aguirre Tagliaferro}, T. and {Biviano}, A. and {De Lucia}, G. and {Munari}, E. and {Garcia Lambas}, D.},
        title = "{Dynamical analysis of clusters of galaxies from cosmological simulations}",
      journal = {\aap},
         year = 2021,
        month = aug,
       volume = {652},
          eid = {A90},
        pages = {A90},
          doi = {10.1051/0004-6361/202140326},
archivePrefix = {arXiv},
       eprint = {2105.09126},
 primaryClass = {astro-ph.CO},
       adsurl = {https://ui.adsabs.harvard.edu/abs/2021A&A...652A..90A}
}

@ARTICLE{Read_2021_MAMPOSStvalidationNbodysim,
       author = {{Read}, J.~I. and {Mamon}, G.~A. and {Vasiliev}, E. and {Watkins}, L.~L. and {Walker}, M.~G. and {Pe{\~n}arrubia}, J. and {Wilkinson}, M. and {Dehnen}, W. and {Das}, P.},
        title = "{Breaking beta: a comparison of mass modelling methods for spherical systems}",
      journal = {\mnras},
         year = 2021,
        month = feb,
       volume = {501},
       number = {1},
        pages = {978-993},
          doi = {10.1093/mnras/staa3663},
archivePrefix = {arXiv},
       eprint = {2011.09493},
 primaryClass = {astro-ph.GA},
       adsurl = {https://ui.adsabs.harvard.edu/abs/2021MNRAS.501..978R}
}

@ARTICLE{2016MNRAS.456.2486D,
       author = {{Despali}, Giulia and {Giocoli}, Carlo and {Angulo}, Raul E. and {Tormen}, Giuseppe and {Sheth}, Ravi K. and {Baso}, Giacomo and {Moscardini}, Lauro},
        title = "{The universality of the virial halo mass function and models for non-universality of other halo definitions}",
      journal = {\mnras},
         year = 2016,
        month = mar,
       volume = {456},
       number = {3},
        pages = {2486-2504},
          doi = {10.1093/mnras/stv2842},
archivePrefix = {arXiv},
       eprint = {1507.05627},
 primaryClass = {astro-ph.CO},
       adsurl = {https://ui.adsabs.harvard.edu/abs/2016MNRAS.456.2486D}
}

@ARTICLE{2016MNRAS.456.2361B,
       author = {{Bocquet}, Sebastian and {Saro}, Alex and {Dolag}, Klaus and {Mohr}, Joseph J.},
        title = "{Halo mass function: baryon impact, fitting formulae, and implications for cluster cosmology}",
      journal = {\mnras},
         year = 2016,
        month = mar,
       volume = {456},
       number = {3},
        pages = {2361-2373},
          doi = {10.1093/mnras/stv2657},
archivePrefix = {arXiv},
       eprint = {1502.07357},
 primaryClass = {astro-ph.CO},
       adsurl = {https://ui.adsabs.harvard.edu/abs/2016MNRAS.456.2361B}
}

@ARTICLE{1915SPAW.......778E,
       author = {{Einstein}, Albert},
        title = "{Zur allgemeinen Relativit{\"a}tstheorie}",
      journal = {Sitzungsberichte der K{\"o}niglich Preussischen Akademie der Wissenschaften},
         year = 1915,
        month = jan,
        pages = {778-786},
       adsurl = {https://ui.adsabs.harvard.edu/abs/1915SPAW.......778E}
}

@ARTICLE{2022A&A...665A.143L,
       author = {{Li}, Pengfei and {McGaugh}, Stacy S. and {Lelli}, Federico and {Schombert}, James M. and {Pawlowski}, Marcel S.},
        title = "{Incorporating baryon-driven contraction of dark matter halos in rotation curve fits}",
      journal = {\aap},
         year = 2022,
        month = sep,
       volume = {665},
          eid = {A143},
        pages = {A143},
          doi = {10.1051/0004-6361/202243916},
archivePrefix = {arXiv},
       eprint = {2208.04326},
 primaryClass = {astro-ph.GA},
       adsurl = {https://ui.adsabs.harvard.edu/abs/2022A&A...665A.143L}
}

@ARTICLE{2026MNRAS.545f2086V,
       author = {{van Daalen}, Marcel P. and {Koutalios}, Ioannis and {Broxterman}, Jeger C. and {Wolfs}, Bart J.~H. and {Helly}, John C. and {Schaller}, Matthieu and {Schaye}, Joop},
        title = "{The resummation model in FLAMINGO: precisely predicting matter power suppression from observed halo baryon fractions}",
      journal = {\mnras},
         year = 2026,
        month = jan,
       volume = {545},
       number = {2},
          eid = {staf2086},
        pages = {staf2086},
          doi = {10.1093/mnras/staf2086},
archivePrefix = {arXiv},
       eprint = {2509.04552},
 primaryClass = {astro-ph.CO},
       adsurl = {https://ui.adsabs.harvard.edu/abs/2026MNRAS.545f2086V}
}

@ARTICLE{2025MNRAS.538.1415S,
       author = {{Sharma}, Divij and {Dai}, Biwei and {Villaescusa-Navarro}, Francisco and {Seljak}, Uro{\v{s}}},
        title = "{A field-level emulator for modelling baryonic effects across hydrodynamic simulations}",
      journal = {\mnras},
         year = 2025,
        month = apr,
       volume = {538},
       number = {3},
        pages = {1415-1426},
          doi = {10.1093/mnras/staf355},
archivePrefix = {arXiv},
       eprint = {2401.15891},
 primaryClass = {astro-ph.CO},
       adsurl = {https://ui.adsabs.harvard.edu/abs/2025MNRAS.538.1415S}
}

@ARTICLE{2019JCAP...03..020S,
       author = {{Schneider}, Aurel and {Teyssier}, Romain and {Stadel}, Joachim and {Chisari}, Nora Elisa and {Le Brun}, Amandine M.~C. and {Amara}, Adam and {Refregier}, Alexandre},
        title = "{Quantifying baryon effects on the matter power spectrum and the weak lensing shear correlation}",
      journal = {\jcap},
         year = 2019,
        month = mar,
       volume = {2019},
       number = {3},
          eid = {020},
        pages = {020},
          doi = {10.1088/1475-7516/2019/03/020},
archivePrefix = {arXiv},
       eprint = {1810.08629},
 primaryClass = {astro-ph.CO},
       adsurl = {https://ui.adsabs.harvard.edu/abs/2019JCAP...03..020S}
}

@ARTICLE{2015JCAP...12..049S,
       author = {{Schneider}, Aurel and {Teyssier}, Romain},
        title = "{A new method to quantify the effects of baryons on the matter power spectrum}",
      journal = {\jcap},
         year = 2015,
        month = dec,
       volume = {2015},
       number = {12},
        pages = {049-049},
          doi = {10.1088/1475-7516/2015/12/049},
archivePrefix = {arXiv},
       eprint = {1510.06034},
 primaryClass = {astro-ph.CO},
       adsurl = {https://ui.adsabs.harvard.edu/abs/2015JCAP...12..049S}
}

@ARTICLE{2013PhRvD..88d3013Z,
       author = {{Zhao}, HongSheng and {Peacock}, John A. and {Li}, Baojiu},
        title = "{Testing gravity theories via transverse Doppler and gravitational redshifts in galaxy clusters}",
      journal = {\prd},
         year = 2013,
        month = aug,
       volume = {88},
       number = {4},
          eid = {043013},
        pages = {043013},
          doi = {10.1103/PhysRevD.88.043013},
archivePrefix = {arXiv},
       eprint = {1206.5032},
 primaryClass = {astro-ph.CO},
       adsurl = {https://ui.adsabs.harvard.edu/abs/2013PhRvD..88d3013Z}
}

@ARTICLE{2018ApJS..239...35D,
       author = {{Diemer}, Benedikt},
        title = "{COLOSSUS: A Python Toolkit for Cosmology, Large-scale Structure, and Dark Matter Halos}",
      journal = {\apjs},
         year = 2018,
        month = dec,
       volume = {239},
       number = {2},
          eid = {35},
        pages = {35},
          doi = {10.3847/1538-4365/aaee8c},
archivePrefix = {arXiv},
       eprint = {1712.04512},
 primaryClass = {astro-ph.CO},
       adsurl = {https://ui.adsabs.harvard.edu/abs/2018ApJS..239...35D}
}

@ARTICLE{2026A&A...707A.153B,
       author = {{Biviano}, A. and {Maraboli}, E.~A. and {Pizzuti}, L. and {Rosati}, P. and {Mercurio}, A. and {De Lucia}, G. and {Ragone-Figueroa}, C. and {Grillo}, C. and {Granato}, G.~L. and {Girardi}, M. and {Sartoris}, B. and {Annunziatella}, M.},
        title = "{CLASH-VLT: The variance in the velocity anisotropy profiles of galaxy clusters}",
      journal = {\aap},
         year = 2026,
        month = mar,
       volume = {707},
          eid = {A153},
        pages = {A153},
          doi = {10.1051/0004-6361/202555439},
archivePrefix = {arXiv},
       eprint = {2508.05195},
 primaryClass = {astro-ph.CO},
       adsurl = {https://ui.adsabs.harvard.edu/abs/2026A&A...707A.153B}
}

@ARTICLE{2025ApJ...987...70A,
       author = {{Abdullah}, Mohamed H. and {Mabrouk}, Raouf H. and {Ishiyama}, Tomoaki and {Wilson}, Gillian and {Amin}, Magdy Y. and {Khattab}, Elamira Hend and {Abdel Rahman}, H.~I.},
        title = "{Quantifying the Velocity Anisotropy Profile of Galaxy Clusters Using the Uchuu Cosmological Simulation}",
      journal = {\apj},
         year = 2025,
        month = jul,
       volume = {987},
       number = {1},
          eid = {70},
        pages = {70},
          doi = {10.3847/1538-4357/adde4b},
archivePrefix = {arXiv},
       eprint = {2504.04575},
 primaryClass = {astro-ph.CO},
       adsurl = {https://ui.adsabs.harvard.edu/abs/2025ApJ...987...70A}
}

@ARTICLE{2023A&A...671A..57S,
       author = {{Seppi}, R. and {Comparat}, J. and {Nandra}, K. and {Dolag}, K. and {Biffi}, V. and {Bulbul}, E. and {Liu}, A. and {Ghirardini}, V. and {Ider-Chitham}, J.},
        title = "{Offset between X-ray and optical centers in clusters of galaxies: Connecting eROSITA data with simulations}",
      journal = {\aap},
         year = 2023,
        month = mar,
       volume = {671},
          eid = {A57},
        pages = {A57},
          doi = {10.1051/0004-6361/202245138},
archivePrefix = {arXiv},
       eprint = {2212.10107},
 primaryClass = {astro-ph.CO},
       adsurl = {https://ui.adsabs.harvard.edu/abs/2023A&A...671A..57S}
}

@ARTICLE{2026arXiv260215934M,
       author = {{Maraboli}, E. and {Biviano}, A. and {Grillo}, C. and {Mercurio}, A. and {Pizzuti}, L. and {Rosati}, P. and {D'Addona}, M.},
        title = "{CLASH-VLT velocity anisotropy profiles in a stack of massive galaxy clusters}",
      journal = {arXiv e-prints},
         year = 2026,
        month = feb,
          eid = {arXiv:2602.15934},
        pages = {arXiv:2602.15934},
          doi = {10.48550/arXiv.2602.15934},
archivePrefix = {arXiv},
       eprint = {2602.15934},
 primaryClass = {astro-ph.GA},
       adsurl = {https://ui.adsabs.harvard.edu/abs/2026arXiv260215934M}
}

@ARTICLE{2017PhRvL.119p1101A,
       author = {{Abbott}, B.~P. and {Abbott}, R. and {Abbott}, T.~D. and {Acernese}, F. and {Ackley}, K. and {Adams}, C. and {Adams}, T. and {Addesso}, P. and {Adhikari}, R.~X. and {Adya}, V.~B. and {Affeldt}, C. and {Afrough}, M. and {Agarwal}, B. and {Agathos}, M. and {Agatsuma}, K. and {Aggarwal}, N. and {Aguiar}, O.~D. and {Aiello}, L. and {Ain}, A. and {Ajith}, P. and {Allen}, B. and {Allen}, G. and {Allocca}, A. and {Altin}, P.~A. and {Amato}, A. and {Ananyeva}, A. and {Anderson}, S.~B. and {Anderson}, W.~G. and {Angelova}, S.~V. and {Antier}, S. and {Appert}, S. and {Arai}, K. and {Araya}, M.~C. and {Areeda}, J.~S. and {Arnaud}, N. and {Arun}, K.~G. and {Ascenzi}, S. and {Ashton}, G. and {Ast}, M. and {Aston}, S.~M. and {Astone}, P. and {Atallah}, D.~V. and {Aufmuth}, P. and {Aulbert}, C. and {AultONeal}, K. and {Austin}, C. and {Avila-Alvarez}, A. and {Babak}, S. and {Bacon}, P. and {Bader}, M.~K.~M. and {Bae}, S. and {Bailes}, M. and {Baker}, P.~T. and {Baldaccini}, F. and {Ballardin}, G. and {Ballmer}, S.~W. and {Banagiri}, S. and {Barayoga}, J.~C. and {Barclay}, S.~E. and {Barish}, B.~C. and {Barker}, D. and {Barkett}, K. and {Barone}, F. and {Barr}, B. and {Barsotti}, L. and {Barsuglia}, M. and {Barta}, D. and {Barthelmy}, S.~D. and {Bartlett}, J. and {Bartos}, I. and {Bassiri}, R. and {Basti}, A. and {Batch}, J.~C. and {Bawaj}, M. and {Bayley}, J.~C. and {Bazzan}, M. and {B{\'e}csy}, B. and {Beer}, C. and {Bejger}, M. and {Belahcene}, I. and {Bell}, A.~S. and {Berger}, B.~K. and {Bergmann}, G. and {Bernuzzi}, S. and {Bero}, J.~J. and {Berry}, C.~P.~L. and {Bersanetti}, D. and {Bertolini}, A. and {Betzwieser}, J. and {Bhagwat}, S. and {Bhandare}, R. and {Bilenko}, I.~A. and {Billingsley}, G. and {Billman}, C.~R. and {Birch}, J. and {Birney}, R. and {Birnholtz}, O. and {Biscans}, S. and {Biscoveanu}, S. and {Bisht}, A. and {Bitossi}, M. and {Biwer}, C. and {Bizouard}, M.~A. and {Blackburn}, J.~K. and {Blackman}, J. and {Blair}, C.~D. and {Blair}, D.~G. and {Blair}, R.~M. and {Bloemen}, S. and {Bock}, O. and {Bode}, N. and {Boer}, M. and {Bogaert}, G. and {Bohe}, A. and {Bondu}, F. and {Bonilla}, E. and {Bonnand}, R. and {Boom}, B.~A. and {Bork}, R. and {Boschi}, V. and {Bose}, S. and {Bossie}, K. and {Bouffanais}, Y. and {Bozzi}, A. and {Bradaschia}, C. and {Brady}, P.~R. and {Branchesi}, M. and {Brau}, J.~E. and {Briant}, T. and {Brillet}, A. and {Brinkmann}, M. and {Brisson}, V. and {Brockill}, P. and {Broida}, J.~E. and {Brooks}, A.~F. and {Brown}, D.~A. and {Brown}, D.~D. and {Brunett}, S. and {Buchanan}, C.~C. and {Buikema}, A. and {Bulik}, T. and {Bulten}, H.~J. and {Buonanno}, A. and {Buskulic}, D. and {Buy}, C. and {Byer}, R.~L. and {Cabero}, M. and {Cadonati}, L. and {Cagnoli}, G. and {Cahillane}, C. and {Calder{\'o}n Bustillo}, J. and {Callister}, T.~A. and {Calloni}, E. and {Camp}, J.~B. and {Canepa}, M. and {Canizares}, P. and {Cannon}, K.~C. and {Cao}, H. and {Cao}, J. and {Capano}, C.~D. and {Capocasa}, E. and {Carbognani}, F. and {Caride}, S. and {Carney}, M.~F. and {Carullo}, G. and {Casanueva Diaz}, J. and {Casentini}, C. and {Caudill}, S. and {Cavagli{\`a}}, M. and {Cavalier}, F. and {Cavalieri}, R. and {Cella}, G. and {Cepeda}, C.~B. and {Cerd{\'a}-Dur{\'a}n}, P. and {Cerretani}, G. and {Cesarini}, E. and {Chamberlin}, S.~J. and {Chan}, M. and {Chao}, S. and {Charlton}, P. and {Chase}, E. and {Chassande-Mottin}, E. and {Chatterjee}, D. and {Chatziioannou}, K. and {Cheeseboro}, B.~D. and {Chen}, H.~Y. and {Chen}, X. and {Chen}, Y. and {Cheng}, H.-P. and {Chia}, H. and {Chincarini}, A. and {Chiummo}, A. and {Chmiel}, T. and {Cho}, H.~S. and {Cho}, M. and {Chow}, J.~H. and {Christensen}, N. and {Chu}, Q. and {Chua}, A.~J.~K. and {Chua}, S.},
        title = "{GW170817: Observation of Gravitational Waves from a Binary Neutron Star Inspiral}",
      journal = {\prl},
         year = 2017,
        month = oct,
       volume = {119},
       number = {16},
          eid = {161101},
        pages = {161101},
          doi = {10.1103/PhysRevLett.119.161101},
archivePrefix = {arXiv},
       eprint = {1710.05832},
 primaryClass = {gr-qc},
       adsurl = {https://ui.adsabs.harvard.edu/abs/2017PhRvL.119p1101A}
}

@ARTICLE{2019A&A...627A..23E,
       author = {{Euclid Collaboration: Adam}, R. and {Vannier}, M. and {Maurogordato}, S. and {Biviano}, A. and {Adami}, C. and {Ascaso}, B. and {Bellagamba}, F. and {Benoist}, C. and {Cappi}, A. and {D{\'\i}az-S{\'a}nchez}, A. and {Durret}, F. and {Farrens}, S. and {Gonzalez}, A.~H. and {Iovino}, A. and {Licitra}, R. and {Maturi}, M. and {Mei}, S. and {Merson}, A. and {Munari}, E. and {Pell{\'o}}, R. and {Ricci}, M. and {Rocci}, P.~F. and {Roncarelli}, M. and {Sarron}, F. and {Amoura}, Y. and {Andreon}, S. and {Apostolakos}, N. and {Arnaud}, M. and {Bardelli}, S. and {Bartlett}, J. and {Baugh}, C.~M. and {Borgani}, S. and {Brodwin}, M. and {Castander}, F. and {Castignani}, G. and {Cucciati}, O. and {De Lucia}, G. and {Dubath}, P. and {Fosalba}, P. and {Giocoli}, C. and {Hoekstra}, H. and {Mamon}, G.~A. and {Melin}, J.~B. and {Moscardini}, L. and {Paltani}, S. and {Radovich}, M. and {Sartoris}, B. and {Schultheis}, M. and {Sereno}, M. and {Weller}, J. and {Burigana}, C. and {Carvalho}, C.~S. and {Corcione}, L. and {Kurki-Suonio}, H. and {Lilje}, P.~B. and {Sirri}, G. and {Toledo-Moreo}, R. and {Zamorani}, G.},
        title = "{Euclid preparation. III. Galaxy cluster detection in the wide photometric survey, performance and algorithm selection}",
      journal = {\aap},
         year = 2019,
        month = jul,
       volume = {627},
          eid = {A23},
        pages = {A23},
          doi = {10.1051/0004-6361/201935088},
archivePrefix = {arXiv},
       eprint = {1906.04707},
 primaryClass = {astro-ph.CO},
       adsurl = {https://ui.adsabs.harvard.edu/abs/2019A&A...627A..23E}
}

@ARTICLE{2024MNRAS.533.3647M,
       author = {{Mitra}, Kaustav and {van den Bosch}, Frank C. and {Lange}, Johannes U.},
        title = "{BASILISK II. Improved constraints on the galaxy-halo connection from satellite kinematics in SDSS}",
      journal = {\mnras},
         year = 2024,
        month = sep,
       volume = {533},
       number = {3},
        pages = {3647-3675},
          doi = {10.1093/mnras/stae2030},
archivePrefix = {arXiv},
       eprint = {2409.03105},
 primaryClass = {astro-ph.CO},
       adsurl = {https://ui.adsabs.harvard.edu/abs/2024MNRAS.533.3647M}
}

@ARTICLE{Pizzuti_2025_refractedgravity,
       author = {{Pizzuti}, L. and {Fantoccoli}, F. and {Broccolato}, V. and {Biviano}, A. and {Diaferio}, A.},
        title = "{Testing refracted gravity with the kinematics of galaxy clusters}",
      journal = {\aap},
         year = 2025,
        month = jun,
       volume = {698},
          eid = {A83},
        pages = {A83},
          doi = {10.1051/0004-6361/202452739},
archivePrefix = {arXiv},
       eprint = {2410.19698},
 primaryClass = {astro-ph.CO},
       adsurl = {https://ui.adsabs.harvard.edu/abs/2025A&A...698A..83P}
}

@ARTICLE{Bonamigo_2018_RXCJ2248_M0416_M1206,
       author = {{Bonamigo}, M. and {Grillo}, C. and {Ettori}, S. and {Caminha}, G.~B. and {Rosati}, P. and {Mercurio}, A. and {Munari}, E. and {Annunziatella}, M. and {Balestra}, I. and {Lombardi}, M.},
        title = "{Dissection of the Collisional and Collisionless Mass Components in a Mini Sample of CLASH and HFF Massive Galaxy Clusters at z {\ensuremath{\approx}} 0.4}",
      journal = {\apj},
         year = 2018,
        month = sep,
       volume = {864},
       number = {1},
          eid = {98},
        pages = {98},
          doi = {10.3847/1538-4357/aad4a7},
archivePrefix = {arXiv},
       eprint = {1807.10286},
 primaryClass = {astro-ph.GA},
       adsurl = {https://ui.adsabs.harvard.edu/abs/2018ApJ...864...98B}
}

@ARTICLE{Bergamini_2023_A2744SL,
       author = {{Bergamini}, Pietro and {Acebron}, Ana and {Grillo}, Claudio and {Rosati}, Piero and {Caminha}, Gabriel Bartosch and {Mercurio}, Amata and {Vanzella}, Eros and {Mason}, Charlotte and {Treu}, Tommaso and {Angora}, Giuseppe and {Brammer}, Gabriel B. and {Meneghetti}, Massimo and {Nonino}, Mario and {Boyett}, Kristan and {Brada{\v{c}}}, Maru{\v{s}}a and {Castellano}, Marco and {Fontana}, Adriano and {Morishita}, Takahiro and {Paris}, Diego and {Prieto-Lyon}, Gonzalo and {Roberts-Borsani}, Guido and {Roy}, Namrata and {Santini}, Paola and {Vulcani}, Benedetta and {Wang}, Xin and {Yang}, Lilan},
        title = "{The GLASS-JWST Early Release Science Program. III. Strong-lensing Model of Abell 2744 and Its Infalling Regions}",
      journal = {\apj},
         year = 2023,
        month = jul,
       volume = {952},
       number = {1},
          eid = {84},
        pages = {84},
          doi = {10.3847/1538-4357/acd643},
archivePrefix = {arXiv},
       eprint = {2303.10210},
 primaryClass = {astro-ph.GA},
       adsurl = {https://ui.adsabs.harvard.edu/abs/2023ApJ...952...84B}
}

@ARTICLE{Maraboli_2025_virialquantitiesfromSL,
       author = {{Maraboli}, Enrico and {Grillo}, Claudio and {Bergamini}, Pietro and {Giocoli}, Carlo},
        title = "{Virial quantities of galaxy clusters from extrapolating strong-lensing mass profiles}",
      journal = {\aap},
         year = 2025,
        month = jun,
       volume = {698},
          eid = {A272},
        pages = {A272},
          doi = {10.1051/0004-6361/202554495},
archivePrefix = {arXiv},
       eprint = {2505.07945},
 primaryClass = {astro-ph.CO},
       adsurl = {https://ui.adsabs.harvard.edu/abs/2025A&A...698A.272M}
}

@ARTICLE{2019MNRAS.484.1598B,
       author = {{Bellagamba}, Fabio and {Sereno}, Mauro and {Roncarelli}, Mauro and {Maturi}, Matteo and {Radovich}, Mario and {Bardelli}, Sandro and {Puddu}, Emanuella and {Moscardini}, Lauro and {Getman}, Fedor and {Hildebrandt}, Hendrik and {Napolitano}, Nicola},
        title = "{AMICO galaxy clusters in KiDS-DR3: weak lensing mass calibration}",
      journal = {\mnras},
         year = 2019,
        month = apr,
       volume = {484},
       number = {2},
        pages = {1598-1615},
          doi = {10.1093/mnras/stz090},
archivePrefix = {arXiv},
       eprint = {1810.02827},
 primaryClass = {astro-ph.CO},
       adsurl = {https://ui.adsabs.harvard.edu/abs/2019MNRAS.484.1598B}
}

@ARTICLE{2007arXiv0709.1159J,
       author = {{Johnston}, David E. and {Sheldon}, Erin S. and {Wechsler}, Risa H. and {Rozo}, Eduardo and {Koester}, Benjamin P. and {Frieman}, Joshua A. and {McKay}, Timothy A. and {Evrard}, August E. and {Becker}, Matthew R. and {Annis}, James},
        title = "{Cross-correlation Weak Lensing of SDSS galaxy Clusters II: Cluster Density Profiles and the Mass--Richness Relation}",
      journal = {arXiv e-prints},
         year = 2007,
        month = sep,
          eid = {arXiv:0709.1159},
        pages = {arXiv:0709.1159},
          doi = {10.48550/arXiv.0709.1159},
archivePrefix = {arXiv},
       eprint = {0709.1159},
 primaryClass = {astro-ph},
       adsurl = {https://ui.adsabs.harvard.edu/abs/2007arXiv0709.1159J}
}

@ARTICLE{2015MNRAS.452.3529V,
       author = {{Viola}, M. and {Cacciato}, M. and {Brouwer}, M. and {Kuijken}, K. and {Hoekstra}, H. and {Norberg}, P. and {Robotham}, A.~S.~G. and {van Uitert}, E. and {Alpaslan}, M. and {Baldry}, I.~K. and {Choi}, A. and {de Jong}, J.~T.~A. and {Driver}, S.~P. and {Erben}, T. and {Grado}, A. and {Graham}, Alister W. and {Heymans}, C. and {Hildebrandt}, H. and {Hopkins}, A.~M. and {Irisarri}, N. and {Joachimi}, B. and {Loveday}, J. and {Miller}, L. and {Nakajima}, R. and {Schneider}, P. and {Sif{\'o}n}, C. and {Verdoes Kleijn}, G.},
        title = "{Dark matter halo properties of GAMA galaxy groups from 100 square degrees of KiDS weak lensing data}",
      journal = {\mnras},
         year = 2015,
        month = oct,
       volume = {452},
       number = {4},
        pages = {3529-3550},
          doi = {10.1093/mnras/stv1447},
archivePrefix = {arXiv},
       eprint = {1507.00735},
 primaryClass = {astro-ph.GA},
       adsurl = {https://ui.adsabs.harvard.edu/abs/2015MNRAS.452.3529V}
}

@ARTICLE{2003ApJ...584..702H,
       author = {{Hu}, Wayne and {Kravtsov}, Andrey V.},
        title = "{Sample Variance Considerations for Cluster Surveys}",
      journal = {\apj},
         year = 2003,
        month = feb,
       volume = {584},
       number = {2},
        pages = {702-715},
          doi = {10.1086/345846},
archivePrefix = {arXiv},
       eprint = {astro-ph/0203169},
 primaryClass = {astro-ph},
       adsurl = {https://ui.adsabs.harvard.edu/abs/2003ApJ...584..702H}
}

@ARTICLE{2015MNRAS.450.3665S,
       author = {{Sereno}, Mauro},
        title = "{CoMaLit - III. Literature catalogues of weak lensing clusters of galaxies (LC$^{2}$)}",
      journal = {\mnras},
         year = 2015,
        month = jul,
       volume = {450},
       number = {4},
        pages = {3665-3674},
          doi = {10.1093/mnras/stu2505},
archivePrefix = {arXiv},
       eprint = {1409.5435},
 primaryClass = {astro-ph.CO},
       adsurl = {https://ui.adsabs.harvard.edu/abs/2015MNRAS.450.3665S}
}

@ARTICLE{2013A&A...558A...1B,
       author = {{Biviano}, A. and {Rosati}, P. and {Balestra}, I. and {Mercurio}, A. and {Girardi}, M. and {Nonino}, M. and {Grillo}, C. and {Scodeggio}, M. and {Lemze}, D. and {Kelson}, D. and {Umetsu}, K. and {Postman}, M. and {Zitrin}, A. and {Czoske}, O. and {Ettori}, S. and {Fritz}, A. and {Lombardi}, M. and {Maier}, C. and {Medezinski}, E. and {Mei}, S. and {Presotto}, V. and {Strazzullo}, V. and {Tozzi}, P. and {Ziegler}, B. and {Annunziatella}, M. and {Bartelmann}, M. and {Benitez}, N. and {Bradley}, L. and {Brescia}, M. and {Broadhurst}, T. and {Coe}, D. and {Demarco}, R. and {Donahue}, M. and {Ford}, H. and {Gobat}, R. and {Graves}, G. and {Koekemoer}, A. and {Kuchner}, U. and {Melchior}, P. and {Meneghetti}, M. and {Merten}, J. and {Moustakas}, L. and {Munari}, E. and {Reg{\H{o}}s}, E. and {Sartoris}, B. and {Seitz}, S. and {Zheng}, W.},
        title = "{CLASH-VLT: The mass, velocity-anisotropy, and pseudo-phase-space density profiles of the z = 0.44 galaxy cluster MACS J1206.2-0847}",
      journal = {\aap},
         year = 2013,
        month = oct,
       volume = {558},
          eid = {A1},
        pages = {A1},
          doi = {10.1051/0004-6361/201321955},
archivePrefix = {arXiv},
       eprint = {1307.5867},
 primaryClass = {astro-ph.CO},
       adsurl = {https://ui.adsabs.harvard.edu/abs/2013A&A...558A...1B}
}

@article{10.1093/mnras/stac1528,
    author = {Mitchell, Myles A and Arnold, Christian and Li, Baojiu},
    title = {A general framework to test gravity using galaxy clusters – VI. Realistic galaxy formation simulations to study clusters in modified gravity},
    journal = {Monthly Notices of the Royal Astronomical Society},
    volume = {514},
    number = {3},
    pages = {3349-3365},
    year = {2022},
    month = {06},
    issn = {0035-8711},
    doi = {10.1093/mnras/stac1528},
    url = {https://doi.org/10.1093/mnras/stac1528},
    eprint = {https://academic.oup.com/mnras/article-pdf/514/3/3349/44234741/stac1528.pdf},
}

@ARTICLE{2025arXiv251214636A,
       author = {{Ahad}, Syeda Lammim and {Reid}, Rashaad and {Mpetha}, Charlie T. and {Taylor}, James E. and {Hildebrandt}, Hendrik and {Hudson}, Michael J. and {Chambers}, Kenneth C. and {de Boer}, Thomas and {Guerrini}, Sacha and {Guinot}, Axel and {Gwyn}, Stephen and {Kilbinger}, Martin and {Van Waerbeke}, Ludovic},
        title = "{Cluster properties as a function of dynamical state in the DESI Legacy x UNIONS surveys}",
      journal = {arXiv e-prints},
         year = 2025,
        month = dec,
          eid = {arXiv:2512.14636},
        pages = {arXiv:2512.14636},
          doi = {10.48550/arXiv.2512.14636},
archivePrefix = {arXiv},
       eprint = {2512.14636},
 primaryClass = {astro-ph.GA},
       adsurl = {https://ui.adsabs.harvard.edu/abs/2025arXiv251214636A}
}

@ARTICLE{2021MNRAS.503..669M,
       author = {{Mpetha}, C.~T. and {Collins}, C.~A. and {Clerc}, N. and {Finoguenov}, A. and {Peacock}, J.~A. and {Comparat}, J. and {Schneider}, D. and {Capasso}, R. and {Damsted}, S. and {Furnell}, K. and {Merloni}, A. and {Padilla}, N.~D. and {Saro}, A.},
        title = "{Gravitational redshifting of galaxies in the SPIDERS cluster catalogue}",
      journal = {\mnras},
         year = 2021,
        month = may,
       volume = {503},
       number = {1},
        pages = {669-678},
          doi = {10.1093/mnras/stab544},
archivePrefix = {arXiv},
       eprint = {2102.11156},
 primaryClass = {astro-ph.CO},
       adsurl = {https://ui.adsabs.harvard.edu/abs/2021MNRAS.503..669M}
}

@ARTICLE{2016A&C....14...35M,
       author = {{Marulli}, F. and {Veropalumbo}, A. and {Moresco}, M.},
        title = "{CosmoBolognaLib: C++ libraries for cosmological calculations}",
      journal = {Astronomy and Computing},
         year = 2016,
        month = jan,
       volume = {14},
        pages = {35-42},
          doi = {10.1016/j.ascom.2016.01.005},
archivePrefix = {arXiv},
       eprint = {1511.00012},
 primaryClass = {astro-ph.CO},
       adsurl = {https://ui.adsabs.harvard.edu/abs/2016A&C....14...35M}
}

@ARTICLE{2013MNRAS.435.1278K,
       author = {{Kaiser}, Nick},
        title = "{Measuring gravitational redshifts in galaxy clusters}",
      journal = {\mnras},
         year = 2013,
        month = oct,
       volume = {435},
       number = {2},
        pages = {1278-1286},
          doi = {10.1093/mnras/stt1370},
archivePrefix = {arXiv},
       eprint = {1303.3663},
 primaryClass = {astro-ph.CO},
       adsurl = {https://ui.adsabs.harvard.edu/abs/2013MNRAS.435.1278K}
}

@ARTICLE{2017MNRAS.468.1981C,
       author = {{Cai}, Yan-Chuan and {Kaiser}, Nick and {Cole}, Shaun and {Frenk}, Carlos},
        title = "{Gravitational redshift and asymmetric redshift-space distortions for stacked clusters}",
      journal = {\mnras},
         year = 2017,
        month = jun,
       volume = {468},
       number = {2},
        pages = {1981-1993},
          doi = {10.1093/mnras/stx469},
archivePrefix = {arXiv},
       eprint = {1609.04864},
 primaryClass = {astro-ph.CO},
       adsurl = {https://ui.adsabs.harvard.edu/abs/2017MNRAS.468.1981C}
}

@ARTICLE{2026arXiv260119861D,
       author = {{Di Dio}, Enea and {Castello}, Sveva and {Bonvin}, Camille},
        title = "{Testing the Equivalence Principle in Galaxy Clusters}",
      journal = {arXiv e-prints},
         year = 2026,
        month = jan,
          eid = {arXiv:2601.19861},
        pages = {arXiv:2601.19861},
          doi = {10.48550/arXiv.2601.19861},
archivePrefix = {arXiv},
       eprint = {2601.19861},
 primaryClass = {astro-ph.CO},
       adsurl = {https://ui.adsabs.harvard.edu/abs/2026arXiv260119861D}
}

@ARTICLE{2019MNRAS.487.1410M,
       author = {{Mitchell}, Myles A. and {Arnold}, Christian and {He}, Jian-hua and {Li}, Baojiu},
        title = "{A general framework to test gravity using galaxy clusters II: A universal model for the halo concentration in f(R) gravity}",
      journal = {\mnras},
         year = 2019,
        month = jul,
       volume = {487},
       number = {1},
        pages = {1410-1425},
          doi = {10.1093/mnras/stz1389},
archivePrefix = {arXiv},
       eprint = {1901.06392},
 primaryClass = {astro-ph.CO},
       adsurl = {https://ui.adsabs.harvard.edu/abs/2019MNRAS.487.1410M}
}

@ARTICLE{2016JCAP...12..024C,
       author = {{Cataneo}, Matteo and {Rapetti}, David and {Lombriser}, Lucas and {Li}, Baojiu},
        title = "{Cluster abundance in chameleon f(R) gravity I: toward an accurate halo mass function prediction}",
      journal = {\jcap},
         year = 2016,
        month = dec,
       volume = {2016},
       number = {12},
          eid = {024},
        pages = {024},
          doi = {10.1088/1475-7516/2016/12/024},
archivePrefix = {arXiv},
       eprint = {1607.08788},
 primaryClass = {astro-ph.CO},
       adsurl = {https://ui.adsabs.harvard.edu/abs/2016JCAP...12..024C}
}

@ARTICLE{2021MNRAS.508.4140M,
       author = {{Mitchell}, Myles A. and {Hern{\'a}ndez-Aguayo}, C{\'e}sar and {Arnold}, Christian and {Li}, Baojiu},
        title = "{A general framework to test gravity using galaxy clusters IV: cluster and halo properties in DGP gravity}",
      journal = {\mnras},
         year = 2021,
        month = dec,
       volume = {508},
       number = {3},
        pages = {4140-4156},
          doi = {10.1093/mnras/stab2817},
archivePrefix = {arXiv},
       eprint = {2106.13815},
 primaryClass = {astro-ph.CO},
       adsurl = {https://ui.adsabs.harvard.edu/abs/2021MNRAS.508.4140M}
}

@ARTICLE{2018MNRAS.477.1133M,
       author = {{Mitchell}, Myles A. and {He}, Jian-hua and {Arnold}, Christian and {Li}, Baojiu},
        title = "{A general framework to test gravity using galaxy clusters - I. Modelling the dynamical mass of haloes in f(R) gravity}",
      journal = {\mnras},
         year = 2018,
        month = jun,
       volume = {477},
       number = {1},
        pages = {1133-1152},
          doi = {10.1093/mnras/sty636},
archivePrefix = {arXiv},
       eprint = {1802.02165},
 primaryClass = {astro-ph.CO},
       adsurl = {https://ui.adsabs.harvard.edu/abs/2018MNRAS.477.1133M}
}

@ARTICLE{2018MNRAS.479.4824H,
       author = {{Hern{\'a}ndez-Aguayo}, C{\'e}sar and {Baugh}, Carlton M. and {Li}, Baojiu},
        title = "{Marked clustering statistics in f(R) gravity cosmologies}",
      journal = {\mnras},
         year = 2018,
        month = oct,
       volume = {479},
       number = {4},
        pages = {4824-4835},
          doi = {10.1093/mnras/sty1822},
archivePrefix = {arXiv},
       eprint = {1801.08880},
 primaryClass = {astro-ph.CO},
       adsurl = {https://ui.adsabs.harvard.edu/abs/2018MNRAS.479.4824H}
}

@ARTICLE{2024A&A...691A.323S,
       author = {{S{\'a}ez-Casares}, I. and {Rasera}, Y. and {Richardson}, T.~R.~G. and {Corasaniti}, P.-S.},
        title = "{The e-MANTIS emulator: Fast and accurate predictions of the halo mass function in f(R)CDM and wCDM cosmologies}",
      journal = {\aap},
         year = 2024,
        month = nov,
       volume = {691},
          eid = {A323},
        pages = {A323},
          doi = {10.1051/0004-6361/202450193},
archivePrefix = {arXiv},
       eprint = {2410.05226},
 primaryClass = {astro-ph.CO},
       adsurl = {https://ui.adsabs.harvard.edu/abs/2024A&A...691A.323S}
}

@ARTICLE{2012MNRAS.425.2128J,
       author = {{Jennings}, Elise and {Baugh}, Carlton M. and {Li}, Baojiu and {Zhao}, Gong-Bo and {Koyama}, Kazuya},
        title = "{Redshift-space distortions in f(R) gravity}",
      journal = {\mnras},
         year = 2012,
        month = sep,
       volume = {425},
       number = {3},
        pages = {2128-2143},
          doi = {10.1111/j.1365-2966.2012.21567.x},
archivePrefix = {arXiv},
       eprint = {1205.2698},
 primaryClass = {astro-ph.CO},
       adsurl = {https://ui.adsabs.harvard.edu/abs/2012MNRAS.425.2128J}
}

@ARTICLE{2017JCAP...03..012V,
       author = {{von Braun-Bates}, F. and {Winther}, H.~A. and {Alonso}, D. and {Devriendt}, J.},
        title = "{The f(Script R) halo mass function in the cosmic web}",
      journal = {\jcap},
         year = 2017,
        month = mar,
       volume = {2017},
       number = {3},
          eid = {012},
        pages = {012},
          doi = {10.1088/1475-7516/2017/03/012},
archivePrefix = {arXiv},
       eprint = {1702.06817},
 primaryClass = {astro-ph.CO},
       adsurl = {https://ui.adsabs.harvard.edu/abs/2017JCAP...03..012V}
}

@ARTICLE{2024JCAP...11..014P,
       author = {{Pizzuti}, Lorenzo and {Boumechta}, Yacer and {Haridasu}, Sandeep and {Pombo}, Alexandre M. and {Dossena}, Sofia and {Butt}, Minahil Adil and {Benetti}, Francesco and {Baccigalupi}, Carlo and {Lapi}, Andrea},
        title = "{Mass modeling and kinematics of galaxy clusters in modified gravity}",
      journal = {\jcap},
         year = 2024,
        month = nov,
       volume = {2024},
       number = {11},
          eid = {014},
        pages = {014},
          doi = {10.1088/1475-7516/2024/11/014},
archivePrefix = {arXiv},
       eprint = {2407.08778},
 primaryClass = {astro-ph.CO},
       adsurl = {https://ui.adsabs.harvard.edu/abs/2024JCAP...11..014P}
}

@ARTICLE{1976ApJ...203..297S,
       author = {{Schechter}, P.},
        title = "{An analytic expression for the luminosity function for galaxies.}",
      journal = {\apj},
         year = 1976,
        month = jan,
       volume = {203},
        pages = {297-306},
          doi = {10.1086/154079},
       adsurl = {https://ui.adsabs.harvard.edu/abs/1976ApJ...203..297S}
}

@ARTICLE{2009MNRAS.399.1106M,
       author = {{Montero-Dorta}, Antonio D. and {Prada}, Francisco},
        title = "{The SDSS DR6 luminosity functions of galaxies}",
      journal = {\mnras},
         year = 2009,
        month = nov,
       volume = {399},
       number = {3},
        pages = {1106-1118},
          doi = {10.1111/j.1365-2966.2009.15197.x},
archivePrefix = {arXiv},
       eprint = {0806.4930},
 primaryClass = {astro-ph},
       adsurl = {https://ui.adsabs.harvard.edu/abs/2009MNRAS.399.1106M}
}

@ARTICLE{2007ApJ...667..760Z,
       author = {{Zheng}, Zheng and {Coil}, Alison L. and {Zehavi}, Idit},
        title = "{Galaxy Evolution from Halo Occupation Distribution Modeling of DEEP2 and SDSS Galaxy Clustering}",
      journal = {\apj},
         year = 2007,
        month = oct,
       volume = {667},
       number = {2},
        pages = {760-779},
          doi = {10.1086/521074},
archivePrefix = {arXiv},
       eprint = {astro-ph/0703457},
 primaryClass = {astro-ph},
       adsurl = {https://ui.adsabs.harvard.edu/abs/2007ApJ...667..760Z}
}

\appendix
\section{Flux-limited surveys}\label{app:SB}

In our main analysis, we assumed that all galaxies within the survey volume are observed. This idealised setting includes only the physical contributions to the signal, namely the gravitational redshift and transverse Doppler effects, and enables us to study observational and modelling systematic effects in isolation. Real surveys, however, are flux-limited, so the observed catalogues include only galaxies brighter than a given apparent-magnitude threshold, $m_{\rm lim}$, which corresponds to a redshift-dependent absolute magnitude limit, $M_{\rm lim}$. Galaxies moving towards the observer are relativistically beamed and therefore appear brighter, making them preferentially selected in flux-limited samples. This selection bias gives rise to the surface-brightness modulation effect \citep{2013MNRAS.435.1278K}. In this section, we investigate observational effects that can occur in a survey combined and therefore incorporate the SB term which was omitted from the total signal in the main analysis.

The change in the apparent luminosity, $l$, of galaxy due to relativistic beaming is given by
\begin{equation}
    \frac{\Delta l}{l}=[3+\alpha(z)]\frac{v_{\rm pec}}{c}\,
    \label{eq:frac_lum}
\end{equation}
where $\alpha$ is the spectral index at the cosmological redshift, $z$, of the source. The number density of objects that pass the detection threshold is
\begin{equation}
    \delta(z)=\frac{d\log{n(<M_{\rm lim}(z))}}{d\log{M}}\,,
\end{equation}
where $n$ indicates number density and $M$ indicates absolute magnitude. Following the assumption of $\alpha=2$ by \citet{2013MNRAS.435.1278K} across our redshift range, the stacked contribution of the effect to the signal can be formulated as \citep{2023A&A...669A..29R}
\begin{equation}
  \bar{\Delta}_{\rm SB}=-\frac{10}{3}\langle \delta(z) \rangle \bar{\Delta}_{\rm TD}\,
\end{equation}
where the TD component is defined in Eq. \eqref{eq:Delta_stack_mass} and the angle brackets denote average over redshift
\begin{equation}
    \langle \delta(z) \rangle = \frac{\int_{z_{\rm min}}^{z_{\rm max}}\delta(z)(dN/dz)dz}{\int_{z_{\rm min}}^{z_{\rm max}}(dN/dz)dz}\,,
\end{equation}
where $dN/dz$ denotes the redshift distribution of galaxies entering the stacked signal, i.e. after all selection effects such as flux limits, redshift cuts, and completeness have been applied.

In order to mimic this effect in our mocks, we inject magnitudes. We start by drawing magnitudes from a \citet{1976ApJ...203..297S} luminosity function configured to follow the r-band, such that $M_{*}=-20.7+5\log_{10}{h}$ and $\alpha_{*}=-1.26$ \citep{2009MNRAS.399.1106M}. We convert the absolute magnitudes to apparent magnitudes following Eq. \eqref{eq:frac_lum}. We adopt $m_{\rm lim}=23$ mag. We found that a $z<1$ spectroscopic survey provides sufficient volume coverage. At the corresponding pivot redshift in such a setting ($z_{\rm piv}=0.66$) this corresponds to $M_{\rm lim}=-20$ mag for our baseline configuration. Following \citet[Table 1,][]{2007ApJ...667..760Z} we adjust our catalogue number density to produce around $8\times10^{6}$ galaxies in the range $0<z<2$, in order to mimic observed number densities. Imposing the apparent magnitude limit yields a total negative bias in the set of radial peculiar velocities surviving the cut. 



\bsp	
\label{lastpage}
\end{document}